\documentclass[9pt]{article}
\usepackage{amsmath}
\usepackage{amsfonts}
\usepackage{epsfig}
\usepackage{color}

\input epsf

\begin{document}
\title{Membrane-protein interactions in mechanosensitive channels}
\author{Paul Wiggins\footnote{Caltech, Department of Physics}  
\& Rob Phillips\footnote{Caltech, Division of Engineering and Applied Physics}}
%\affil{Caltech, Pasadena, CA 91125}
\maketitle

\abstract
In this paper, we examine the mechanical role of the lipid bilayer in ion channel 
conformation and function with specific reference to the case of the mechanosensitive 
channel of large conductance (MscL).
%, a transmembrane protein believed to play an important 
%role in the osmoregulation of bacteria (Blount and Moe, 1999). 
In a recent paper 
(Wiggins and Phillips, 2004), we argued that mechanotransduction very naturally arises 
from lipid-protein interactions by invoking a simple analytic model of the MscL channel
and the surrounding lipid bilayer. 
In this paper, we focus on improving and expanding this analytic framework for studying
lipid-protein interactions with special attention to MscL. Our goal is to generate simple
scaling relations which can be used to provide qualitative understanding of the role of membrane 
mechanics in protein  function and to quantitatively interpret experimental 
results. For the MscL channel, we find that the free energies induced by lipid-protein 
interaction are of the same order as the free energy differences between conductance states 
measured by Sukharev {\it et al.} (1999). We therefore conclude that the mechanics of the 
bilayer plays an essential role in determining the conformation and function of the channel. 
Finally, we compare the predictions of our model to experimental results 
from the recent investigations of the MscL channel by Perozo {\it et al.} (2002a,b), Powl 
{\it et al.} (2003), Yoshimura {\it et al.} (2004), and others and suggest a suite
of new experiments.
% and 
%discuss the physical principles underlying channel gating and mechanotransduction.  

\section{Introduction}
The mechanosensitive channel of large conductance (MscL) is a compelling example of the 
interaction of a membrane protein and the surrounding lipid bilayer membrane. 
MscL is gated mechanically (Blount and Moe, 1999) and belongs to a growing 
class of proteins which have been determined to be mechanosensitive (Gillespie and Walker, 2001; 
Hamill and Martinac, 2001). 
In a recent short paper (Wiggins and Phillips, 2004), we have argued that the mechanics of the bilayer 
is an important partner in the phenomena of mechanotransduction and channel function. 
In particular, we considered a simplified model where only the radius of the channel changes in 
transitions between the open and closed state. 
In this paper, we present our free energy calculations in more generality and detail. 
Specifically, we have calculated the free energy due to the bilayer deformation  as
a result of the presence of a membrane protein using an analytic model developed for the study of bilayer mechanics 
(Canham, 1970; Helfrich, 1973; Evans, 1974). 
Many of the theoretical techniques exploited here have already been used with success in describing the role 
of the bilayer in the mechanics of the Gramicidin channel (e.g. Huang, 1986). 
In this paper,
we have applied asymptotic approximations to the exact solutions of this model, permitting many of the 
important results to be expressed, estimated, and understood with simple scaling relations. 
These scaling relations are then applied to estimate the relative importance of each and every term 
in the bilayer free energy budget. We find that the bilayer 
deformation free energy can be of the same order 
as the free energy differences between conformational states of the MscL channel as measured by 
Sukharev {\it et al.} (1999). These results strongly suggest that 
bilayer deformation  plays an important 
role in determining the protein conformation, and therefore function, of transmembrane proteins in general, 
and MscL in particular. Although we have explicitly estimated 
the size of the bilayer deformation energy  exclusively for the geometry of MscL, 
the results can easily be re-evaluated and reinterpreted in the context of other transmembrane 
proteins and mechanosensitive channels, in particular: MscS (Bass, 2002) Alamethicin (Opsahl and Webb, 1994), etc.
%We focus on exploring a new facet of the intriguing bilayer-inclusion interaction story: 
%the radial dependence of the bilayer deformation energy. 
%This radial dependence plays an important role in the function of 
%proteins (like MscL) which undergo conformational changes that significantly change the size of the protein 
%bilayer interface. We discuss the ways in which channels and other proteins can harness the 
%bilayer deformation energy  to 
%aid in their function. 
We emphasize that our goal in this current work is not to attempt a 
quantitative understanding of all of the degrees of freedom of the channel and bilayer, but rather to build 
a tractable model for the role of bilayer mechanics in the function of the MscL channel, while developing
the model in more detail than in our previous paper (Wiggins and Phillips, 2004).

The MscL channel is gated by membrane tension and has been studied extensively in patch clamp
experiments (Sukharev {\it et al.}, 1999; Perozo {\it et al.}, 2002a). 
While several substates have been identified (Sukharev {\it et al.},1999), the channel 
typically resides in one of two primary conductance states. At low tension the channel is 
almost exclusively closed (C). As the tension is increased the open state (O) 
becomes ever more prevalent, until it dominates at high tension.
Rees and coworkers have solved the structure for one conformation using X-ray crystallography 
which appears to be the closed state (Chan {\it et al.}, 1998).  The open state has been modeled by a 
number of groups (Sukharev {\it et al.}, 2001; Betanzos {\it et al.}, 2002; Perozo {\it et al.}, 2002b).

The outline of the paper is as follows. In section 2, we briefly discuss the bilayer model, 
then present a table of results which shows the relative importance of
different free energy penalties for bilayer deformation and then define the generalized 
forces we use to discuss the 
effects of bilayer deformation induced by protein conformational change. 
In section 3, we estimate the sizes of the bilayer deformation energy 
and forces for MscL, give a brief physical discussion of mechanisms which give 
rise to the bilayer deformation energy, and discuss the scaling of these bilayer deformation energies.
In section 4, we compare our predictions for a two state MscL model 
to experimental measurements made by Perozo {\it et al.} (2002a,b), Powl {\it et al.} (2003), 
Yoshimura {\it et al.} (2004), and others. In the appendix, we provide a unit conversion table, bilayer 
parameters and full names, detailed derivations, and a discussion of the approximations used.

\section{Free Energy of the Bilayer-Inclusion System}
\label{FEBIS}
We begin by  considering the free energy of the system as a 
whole: protein and bilayer. 
We assume that the system is in thermal equilibrium and define the free energy differences between
states in the standard way:
\begin{equation}
\Delta G_{(i)} \equiv -kT \log \left(\frac{{\cal P}_{i}}{{\cal P}_{C}}\right), 
\end{equation} 
where ${\cal P}_{i}$ is the probability of state $i$ and the free energy differences 
are defined with respect to the closed state. We can divide each of these free energies into
two parts 
\begin{equation}
G = G_{P}+G_{\cal M},
\end{equation} 
where $G_{P}$ is the free energy associated with the protein's conformation and $G_{{\cal M}}$ is the free
energy induced in the bilayer by the protein inclusion and includes both a deformation free energy from the 
bulk of the bilayer and an interaction energy at the interface between the inclusion and bilayer. 
For the sake of brevity we will usually refer to both of these bilayer-related contributions to the free energy 
as the bilayer deformation energy.
While a complete understanding of channel gating and function must encompass knowledge of both components
of the free energy, $G_P$ and $G_{\cal M}$, our analysis in this paper 
centers almost entirely on the bilayer deformation energy, 
$G_{{\cal M}}$. Several groups have used molecular dynamics (MD) and related techniques to study $G_P$
or $G$ in its entirety (Gullingsrud {\it et al.}, 2001; Gullingsrud and Schulten, 2002; Bilston and Mylvaganam, 2002; 
Elmore and Dougherty, 2003; Gullingsrud and Schulten, 2003) 
but as is often the case for biological systems, 
these studies have been handicapped by the size of the MscL system. It is too complex for direct 
simulation on biologically relevant time scales. 
From an experimental standpoint, Sukharev {\it et al.} have measured the free energy differences between 
different states (Sukharev {\it et al.}, 1999) in MscL and have found differences of order $10\, kT$. 
These results reveal the energy scale associated with MscL gating and provide a reference by
which different contributions to the free energy will be judged for their importance. 

Our first goal in what follows is to persuade the reader that $G_{{\cal M}}$ is large enough to be of interest.
That is, since it is clear that there are several distinct contributions to the overall
free energy budget, we illustrate that the contribution due to the inclusion-induced
bilayer deformation is comparable to the measured free energy differences between states. 
Since these contributions to $G_{{\cal M}}$ are of the same order of magnitude as $\Delta
G$, we  conclude that the effects of bilayer deformation are potentially 
interesting. Perozo {\it et al.} (2002a)  have already answered this question experimentally, 
demonstrating that bilayer 
characteristics such as lipid actyl chain length significantly effect the free energy. 

\subsection{The Calculation of the Bilayer Free Energy}
\label{mfe_section} 
The elastic deformation of the bilayer surrounding the channel is approximated with a model developed by 
Canham (1970), Helfrich (1973), and Evans (1974). Huang (1986) 
has applied this model to calculate the deformation energies induced by inclusions. These
calculations have been elaborated upon by others: notably by Andersen and coworkers 
(Nielson {\it et al.}, 1998; Lundb{\ae}k {\it et al.}, 1996; Lundb{\ae}k and Andersen, 1994; 
Goulian {\it et al.}, 1998) and Dan and 
coworkers (Dan {\it et al.}, 1994; Dan {\it et al.}, 1993; Dan and Safran, 1995; Dan and Safran, 1998).
Specifically, Goulian {\it et al.} (1998) have studied a similar model including applied 
tension. The bilayer deformation energy in this model  is given
by
\begin{equation}
G_{\cal M} = G_+ + G_-+G_{\rm I}, 
\end{equation}
where
\begin{equation}
G_\pm = \int_{{\cal M}}d^2\sigma\,
\left({\textstyle \frac{1}{ 4}}K_B[{\rm tr}\ {\bf S}_\pm(\vec{x})\mp C_\pm]^2+\frac{K_G}{2}\det{\bf S}_\pm(\vec{x})+ 
\alpha_\pm\right),
\end{equation}
and
\begin{equation}
G_{\rm I} = \int_{{\cal M}}d^2\sigma\ {\textstyle \frac{1}{ 2}}K_A\left({\textstyle \frac{u}{a}}\right)^2.
\end{equation}
$G_\pm$ are the free energies due to the curvature and the tension 
in the top and bottom surfaces of the bilayer and $G_{\rm I}$ is the interaction 
free energy between these two surfaces. Locally, the curvature of the top (or bottom) surface of
the bilayer is described by the shape operator, ${\bf S}_{\pm}(\vec{x})$, a rank two tensor.
The trace of this tensor is twice the mean curvature and its determinant is the Gaussian curvature.
The energetic cost for increasing the mean curvature of the top (or bottom) surface of the bilayer is the 
bending modulus, $K_B/2$. The energetic cost for increasing the Gaussian curvature of the 
top (or bottom) surface is the Gaussian bending modulus, $K_G/2$. We have chosen this normalization so that 
the effective moduli for the bilayer as a whole are $K_B$ and $K_G$.
 
%For all the calculations we will assume that the bending moduli of 
%layers are identical and the composite moduli for the bilayer as a whole are 
%\begin{eqnarray}
%K_B = 2K_B^\pm, \\
%K_G = 2K_G^\pm. 
%\end{eqnarray}
%We will use a spherical membrane 
%patch as the equilibrium geometry of the membrane in the absence of the inclusion. This will serve
%as the background solution which is then perturbed by the
%presence of the membrane protein. We will assume the ratio of the channel radius, $R$, 
%to the radius of curvature of the patch, $R_0$, is small and we will only keep the 
%lowest order contribution from this background curvature. 
The addition of certain surfactants and non-bilayer lipids  
results in the lowest energy conformation of a single layer of 
lipids being curved (Israelachvili, 1991; Gruner, 1989). 
This spontaneous curvature is introduced into the model through non-zero 
values of $C_\pm$, the spontaneous curvatures of the upper and lower layers of the bilayer, respectively.
We define the composite and the mean spontaneous curvatures for the bilayer as 
\begin{eqnarray}
C &\equiv& {\textstyle\frac{1}{2}}(C_+-C_-), \\
\overline C &\equiv& {\textstyle\frac{1}{2}}(C_++C_-), 
\end{eqnarray}
respectively.
The energetic penalty associated with changes in the area of the top and bottom surfaces of the 
bilayer are the tensions, $\alpha_\pm$, respectively. 
We assume that the tensions in the two layers are equal since, 
on long time scales, the lipids can switch between the two leaflets in order to equalize the tension.
The total tension, $\alpha=2\alpha^\pm$, is an externally tunable parameter. 
(See the appendix \ref{tensionexp} for further discussion.)  
$2u$ is the difference 
between the local thickness of the bilayer and the equilibrium
thickness $2a$. The energetic cost for changing the thickness of the bilayer is the compression-expansion 
modulus, $K_A$. 
For further discussion of the model, the reader is invited to view the extensive discussions 
in the literature (Helfrich, 1973; Huang, 1986; Dan {\it et al.}, 1994; Goulian {\it et al.} 1998; etc.)
\begin{figure}
\begin{center}
\leavevmode 
\epsfig{file=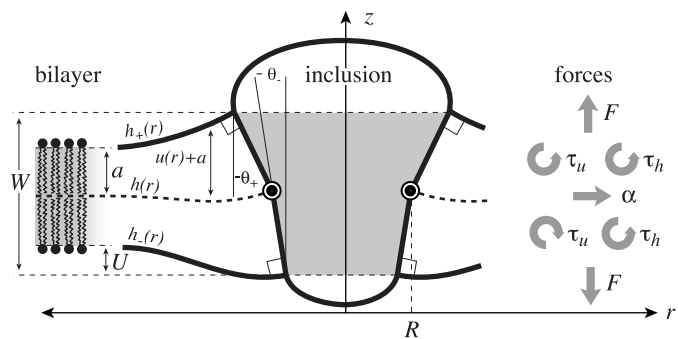}
\caption{\label{basic} A schematic picture of the bilayer-inclusion model.
The geometry of the inclusion is described by four parameters: the radius $R$, the thickness $W$, 
and the radial slopes $H'_\pm$ of the top and bottom surfaces of the bilayer, respectively. If the
surfaces of the bilayer are locally normal to the interface of the inclusion, as depicted above, 
$H'_\pm = \theta_\pm$ in the small angle approximation. The bilayer equilibrium thickness is $2a$.
The fields $h_\pm(r)$ are the $z$ displacements of the top and bottom surfaces
of the bilayer, respectively. Their average is the mid-plane displacement, $h(r)$, 
and half their difference is $u(r)+a$. $u(r)$ is the local thickness deformation of a single leaflet 
of the bilayer. 
At the interface, twice this deformation, $2U$, is the hydrophobic mismatch, $W-2a$. 
The generalized forces on the inclusion induced by the 
bilayer are depicted for positive values. $F$ is the expansion-compression force, $\alpha$ is the 
tension, $\tau_h$ is the mid-plane torque, and $\tau_u$ is the shape torque. }
\end{center}
\end{figure}

The presence of the channel will perturb the bilayer locally. To
calculate the perturbation to the free energy due to the channel, we will assume 
that the radius of curvature corresponding to the vesicle or cell in which the 
inclusion is embedded is very 
large in comparison to the length scale of the inclusion itself and that the 
perturbation due to the bilayer inclusion is small enough to allow the equations 
to be linearized. In this approximation scheme, the bilayer deformation energy  is
\begin{equation}
G_{\cal M} = \int_{\cal M'} d^2x\, {\cal G}, 
\end{equation}
where ${\cal G}$ is the expanded effective free energy density (written out in 
its expanded form in the appendix) and $\cal M'$ is a 
Cartesian plane minus a circular inclusion of radius $R$. We can safely integrate out to
infinity since the perturbation to the free energy density is localized around the inclusion. 
To construct the effective free energy density we describe the out-of-plane
displacements of the upper and lower surfaces of the bilayer with the functions $h_+(\vec{x})$ and $h_-(\vec{x})$, 
respectively, on $\cal M'$, as shown in fig.~\ref{basic}. It is more transparent 
to work with the linear combinations
of these two functions (Fournier, 1999), namely, 
\begin{eqnarray}
h(\vec{x}) &=& {\textstyle\frac{1}{2}}(h_++h_-), \\
u(\vec{x}) &=& {\textstyle\frac{1}{2}}(h_+-h_-)-a,
\end{eqnarray}
where $h$ is the average position of the upper and lower surfaces of 
the bilayer which we will refer to as the mid-plane and $u$ is half the difference of the bilayer
thickness and the equilibrium thickness. The overall structural picture is shown in fig. \ref{basic}
 where the localized perturbation of the bilayer is depicted 
schematically.

A minimization of the effective free energy (Huang, 1986)  gives two decoupled differential 
equations (Fournier, 1999) for the equilibrium 
configuration in the fields $u(\vec{x})$ and $h(\vec{x})$, namely, 
\begin{eqnarray}
0 &=& \left[K_B \nabla^4 - \alpha\nabla^2 +{\textstyle \frac{K_a}{a^2}}\right]u \label{equ1}\\
0 &=& \left[K_B \nabla^2 - \alpha\right]h, \label{equ2}
\end{eqnarray}
which are again discussed at length in the literature (Huang, 1986; Nielsen {\it et al.}, 1998; Fournier, 1999). 
The solution to these equations for the fields $h(\vec{x})$ and $u(\vec{x})$ can be written in terms of modified 
Bessel functions in cylindrical coordinates (Huang, 1986).  

\begin{figure}
\begin{center}
\leavevmode 
\epsfig{file=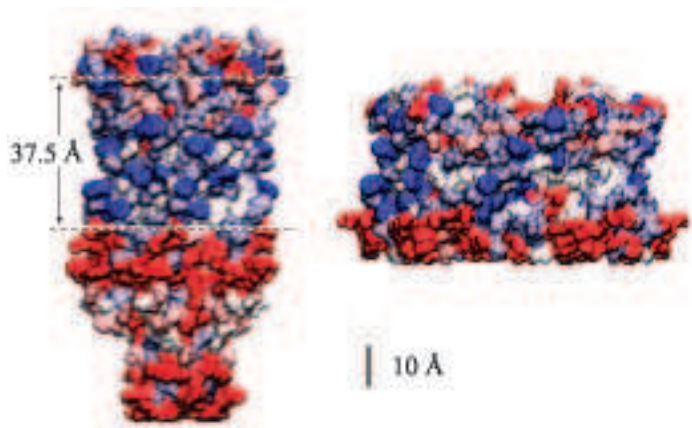}
\caption{\label{rendered} Models of the closed and open states colored by hydrophobicity (Sukharev {\it et al.}, 2001). 
While the general region
spanned by the membrane is evident from the hydrophobic regions on the protein interface, it is difficult to 
precisely define the thickness of this region. A closed states thickness has been inferred from the data of 
Powl {\it et al.} (2003) and this region is schematically marked on the model of the closed state. Additional
confirmation of this estimate for the
hydrophobic thickness comes from the simulation of Elmore and Dougherty (2003). }
\end{center}
\end{figure}
Due to the hydrophobic residues of the protein inclusion, we assume that the bilayer 
adheres to the external surface of the protein.  As will be described in more
detail below, the
matching condition at this surface 
dictates half the boundary conditions for the bilayer 
(the remaining boundary conditions dictate that the bilayer is unperturbed at infinity). We consider
proteins with azimuthal (cylindrical) symmetry. 
While the Mscl channel is not truly azimuthally symmetric, as a homo-pentamer, 
it is highly symmetric, at least in the 
closed state as the X-ray crystallography structure has demonstrated (Chang {\it et al.}, 1998).
In order to clearly distinguish values of the functions at the boundaries from 
the corresponding functions themselves, we will denote these parameters
with capital letters. We fix the bilayer thickness, $2U+2a$, to match the hydrophobic 
thickness of the protein, $W$,
at the interface, $r=R$:
\begin{equation}
u(R) = U = {\textstyle\frac{1}{2}}W-a.
\end{equation}
$2U$ is called the hydrophobic mismatch since it is the difference between the 
equilibrium thickness of the bilayer, $2a$, and the thickness of the protein, $W$. 
For real proteins it is quite difficult to define exactly what one means 
by this region since real structures are not purely hydrophobic in the transmembrane 
region. 
%(???? would like to reference something here.???). 
The closed state and a proposed model
of the open state colored by hydrophobicity are depicted in figure \ref{rendered}.

We also specify the radial
derivatives of $h_\pm$ at the boundary as 
\begin{equation} 
h'_{\pm}(R) = H_\pm'
\end{equation}
or alternatively,
\begin{eqnarray}
H' \equiv h'(R) &=& {\textstyle\frac{1}{2}}(H_+'+H_-'),\\
U' \equiv u'(R) &=& {\textstyle\frac{1}{2}}(H_+'-H_-'),
\end{eqnarray}
where $'$ is the derivative with respect to $r$, the radial distance from the inclusion. 
A physical interpretation of these slopes 
might be to assume the bilayer's surfaces are normal to the protein's surface at the
boundary, although this need not be the case (Nielson {\it et al.}, 1998).
At infinity we assume that the bilayer is unperturbed which may be cast in mathematical terms as
\begin{eqnarray}
h(\infty) &=& 0, \\
u(\infty) &=& 0.
\end{eqnarray}
Solving the equilibrium equations for a given set of 
boundary conditions and plugging these solutions into the surface integral for
bilayer deformation energy  results in the bilayer deformation energy  for a 
given configuration of the protein (Huang, 1986).
Each protein configuration corresponds to a different outcome
for the bilayer deformation energy .  
This energy has been divided into several contributions based on the physical 
mechanism giving rise to it. 
In table \ref{table}, we present a summary of these results. 
Brief derivations may be found in the appendix.
Generally, the bilayer deformation energies lend themselves to simple scaling laws, 
except for two cases: thickness 
and mid-plane deformation. In these cases the exact results to the model are somewhat 
complicated and the results that appear in the table are limits which are 
derived and discussed in the appendix. 
\begin{table}
\begin{center}
\begin{tabular}{|ccc|}
\hline

Physical Mechanism & Energy ($G_{\cal M}$) & for MscL  \\
\hline 
& &  \\ 
Areal Deformation & $G_A = -\alpha \cdot A$ & $10 kT$  
\\ & & \\

Gaussian Curvature    & $G_G = -\pi K_G\left(H'^{\, 2}+U'^{\, 2}\right)$ & $1kT$  
\\ & &  \\

%Background Curvature & $G_{R} = 2K_B\frac{H'}{R_0} \cdot \cal C$ &   $\ll kT $ 
% \\ & &  \\ 
%
Spontaneous Curvature & $G_{C} = K_B \left(C H'+\overline{C}U'
\right)\cdot {\cal C}$ & $10kT$  \\ & &  \\ 

Bilayer Interface  & $G_{\sigma} = \sigma W\cdot {\cal C}$ & $10kT$ 
\\ & &  \\

Mid-Plane Deformation \dag & $G_{H} = {\textstyle \frac{1}{2}}\sqrt{\alpha K_B} H'^{\, 
2}\cdot {\cal C}$ & $<kT$  \\ & &  \\

Thickness Deformation \dag & $G_{U} = {\textstyle \frac{1}{2}}{\cal K}U^2\cdot {\cal C} 
$ &  $10 kT$  \\  & &  \\
\hline
\end{tabular}
\caption{Summary of results for inclusion-induced bilayer free energies. The free energies are written symbolically 
followed by an estimate of the size of the contribution to the nearest order of magnitude for a typical MscL system in  
patch clamp experiments. In the following section more detailed estimates are made.
The free energies have been factored to emphasize their radial dependence. 
Tension-like terms are proportional to the area, $A\equiv \pi R^2$. Line-tension-like 
terms are proportional to the circumference, ${\cal C} \equiv 
2\pi R$. $\cal K$ is a composite elastic constant defined in section \ref{TDT}. 
$\sigma$ is an interface energy discussed in section \ref{IET}. \dag Dominant 
scaling for asymptotic results. 
\label{table}}
\end{center}
\end{table}

\subsection{Connection between $H_\pm'$ and Channel Geometry}
\label{nia}
Recall from the discussion above that the energetics of the composite system of the inclusion
and the bilayer depends in part on the geometric parameters $H_\pm'$ that determine
how the bilayer joins the protein at the interface.
The appropriate bilayer slope boundary condition is still somewhat of an open question. 
Some authors
 have treated these conditions as free, minimizing the bilayer deformation energy  with
respect to them, while others have assumed that the bilayer surfaces are normal to the 
protein surface (see refs. in Nielson {\it et al.}, 1998). 
Most of  our results will be expressed in terms of $H_\pm'$ which is
independent of any particular assumption about these boundary conditions,
 though we will assume the normal interface boundary conditions in our
concrete physical discussions. We will also discuss the free boundaries briefly. 
If we assume that the mid-plane of the lipid bilayer interface is normal
to the protein  and that transmembrane domains M1 and M2 are rigid 
and aligned this dictates that
\begin{eqnarray}
H' &=& H_+' = H_-' \\
U' &=& 0.
\end{eqnarray} 
This can be recast verbally as the statement that the top and bottom surfaces of the bilayer 
have the same slope at the boundary and there is no bend in 
the inclusion interface. In the small angle limit, $H'$ can be
replaced by the angle away from normal of the interface. 
If we do introduce a bend in the middle of the interface, the orientations of the upper and lower 
interfaces are
independent. Assuming that the interface of the bilayer 
is normal to the protein surface, we can replace the slopes with 
the angles away from normal, $\theta_\pm$, (Dan and Safran, 1998) in the small angle
limit, as pictured schematically in fig. \ref{basic}.

\subsection{Forces, Torques, and Tensions}
\label{defforce}
The physical effects of bilayer deformation on the inclusion conformation 
can be recast in a more
intuitive form by appealing to forces, tensions, and torques rather than free energies.  
For example, most of the bilayer deformation energies will generate a 
tension on the interface due to their radial dependence. The applied tension, 
$\alpha$, is not the whole story! 
The generalized forces are obtained by differentiating the bilayer deformation energy 
with respect to bilayer excursions. Implicitly, these generalized forces are defined through
\begin{equation} 
dG_{\cal M} = -\alpha_\Sigma dA -\tau_+ dH_+'-\tau_- dH_-'-FdW, 
\end{equation}
where $A\equiv\pi R^2$ is the area of the protein, $H_\pm'$ are the slopes of the 
bilayer surfaces at the interface, and $W$ is the thickness of the hydrophobic region of the
protein. Explicitly, these generalized forces may be written as 
\begin{eqnarray}
\alpha_\Sigma &\equiv& -\frac{1}{2\pi R}\left(\frac{\partial G_{\cal M}}{\partial R}\right)_{T,W,H_\pm'}, \\
\tau_{\pm} &\equiv& -\left(\frac{\partial G_{\cal M}}{\partial H_\pm'}\right)_{T,A,W,H_\mp'}, \\
{F} &\equiv& -\left(\frac{\partial G_{\cal M}}{\partial W}\right)_{T,A,H_\pm'}. 
\end{eqnarray}
Since we have already used $\alpha$ to denote the applied tension, we use $\alpha_\Sigma$ to denote the net
radial tension on the inclusion interface: the sum of the applied tensions and other bilayer 
deformation induced contributions. When the tension is positive, it is tensile. 
$F$ is the compression-expansion 
force, normal to the plane of the bilayer, acting on the inclusion. When the compression-expansion force is 
positive, it acts to induce inclusion-thickness expansion.  
$\tau_{\pm}$ are cylindrical torques acting on the top and bottom surfaces of the inclusion around the 
mid-plane. 
It will usually be more convenient to work with the torques complimentary to $H'$ and $U'$ rather 
than $H'_\pm.$
We define the mid-plane torque as the cylindrical torque on the interface as a whole:
\begin{equation}
\tau_h \equiv \tau_+ + \tau_- = -\left(\frac{\partial G_{\cal M}}{\partial H'}\right)_{T,A,W,U'}.
\end{equation}
When the mid-plane torque is positive, it acts to induce increases in the mid-plane slope. 
This cylindrical torque is generated by bending stresses alone and is therefore related to the
principal curvatures at the boundary (Landau and Lifshitz, 1986) through the relation
\begin{equation}
\frac{\tau_h}{{\cal C}} = -K_B\left( R_\Vert^{-1} + R_\bot^{-1} -C\right) - K_GR_\Vert^{-1},
\label{bs}
\end{equation}
where $\cal C$ is the circumference of the inclusion, $R_\Vert^{-1}$ and $R_\bot^{-1}$ are the
principal curvatures at the boundary of the midplane, in the directions parallel and perpendicular to the boundary, 
respectively. (For azimuthally symmetric surfaces the principal curvatures are always radial and 
azimuthal, and furthermore the azimuthal curvature is 
$R_\Vert^{-1} = -r^{-1}\sin \theta_N$ where $r$ is the cylindrical radius, and $\theta_N$ is the 
angle of the upward surface normal away from vertical.  For example, see Boal, 2002.) We can define the shape 
torque as the cylindrical torque complimentary to $U'$:
\begin{equation}
\tau_u \equiv \tau_+ - \tau_- = -\left(\frac{\partial G_{\cal M}}{\partial U'}\right)_{T,A,W,H'}.
\end{equation}
When the shape torque is positive, it induces radial expansion at the mid-plane and radial compression at 
the outer surfaces of the bilayer. When the shape torque is negative, it induces radial compression at 
the mid-plane and radial expansion at the outer surfaces of the bilayer. 
The bending stress picture of the shape torque is somewhat more complicated than for the mid-plane torque
due to the interaction between the two layers. The generalized forces are depicted in fig. \ref{basic}
for positive values and their physical interpretation and size are discussed in section \ref{FEEaPI}.

\subsection{Relation between pressure gradients and generalized forces}
Another way to recast the interaction between the membrane protein and the surrounding 
bilayer is by introducing the notion of pressure gradients.
Cantor (1997, 1999)  
has made calculations of the out-of-plane pressure gradients in the bilayer.
He has shown that the pressure is compressive in the middle of the bilayer and expansive near the
surface. Cantor (1997) and de Kruijff (1997) have discussed the effects of this gradient 
on protein conformation. If the $\alpha$-helices (MscL's transmembrane domains M1 and M2) can be 
interpreted (to a first approximation) as rigid, the effects of this pressure gradient are to produce
a tension and cylindrical torques. The tension on the interface is the integrated bilayer pressure,
\begin{equation}
\alpha_P = \int^a_{-a}dz\ P(z),
\end{equation}
where $z$ is the position in the bilayer, running from $-a$ to $a$. This integrated tension 
must be the net tension $\alpha_{\Sigma}$. 
If we allow the inclusion to have a hinge at $z=0$, cylindrical torques about 
this circumference are induced on each section of the inclusion. (See fig \ref{basic}) In the small angle
limit, these cylindrical torques are
\begin{eqnarray}
\tau_{+,P} &=& R\int^{2\pi}_0 d\phi \int^a_0 dz\ P(z)z,  \\
\tau_{-,P} &=& -R\int^{2\pi}_0 d\phi \int^0_{-a} dz\ P(z)z,  
\end{eqnarray}
where the torques have been defined to match our previous definitions
in section \ref{defforce}, when the angles made by two surfaces of the interface, 
$\theta_\pm$, are defined such that
\begin{equation}
\theta_\pm = H'_\pm.
\end{equation}
The $\tau_{\pm,P}$ must correspond to our $\tau_{\pm}$. The effects of the pressure gradient on 
our constrained system are neatly reduced to three of the generalized forces we have already discussed.
The fourth force, $F$, is just the integrated shear stress.

\section{Free Energy Estimates and Physical Interpretation}
\label{FEEaPI}
In section \ref{FEBIS}, we summarized the bilayer model and presented the 
lowest order contributions to the bilayer deformation energy  in table \ref{table}.
The aim of the present section is to revisit each of these individual contributions to the overall free energy, 
estimate its magnitude for MscL,
and discuss the scaling and physical mechanism giving rise to the bilayer deformation energy.
In order to estimate the bilayer deformation energies, we need structural information for MscL. 
From X-ray crystallography data (Chang {\it et al.}, 1998), in the closed state, MscL appears to have an 
external radius of roughly $23\ {\rm \AA}$. Sukharev {\it et al.} (2001) have speculated that the open
state's external radius is roughly $35\ {\rm \AA}$. We use typical bilayer 
elastic parameters as summarized in the appendix. In addition, the appendix contains a brief discussion of the 
scaling of these parameters with bilayer thickness.
Numerical results are multiplied by scaling relations to remind the reader what values have been used 
in their computation and how the free energies scale with changes in inclusion geometry, tension, etc.

\subsection{Areal Deformation }
\begin{figure}
\begin{center}
\epsfig{file=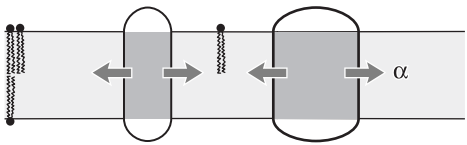}
\caption{\label{area} A cartoon of areal deformation. Tension, represented by
the arrows, is transmitted 
through the bilayer to the inclusion.  
For positive biaxial tension, radial 
expansion of the inclusion reduces the free energy of the bilayer. The 
vesicle or cell can be viewed as a bilayer reservoir where tension is the 
energetic cost per unit area of bilayer in the local system. }
\end{center}
\end{figure}

The areal deformation free energy is the dominant tension-dependent term and typically provides 
the mechanism for opening the MscL channel. The physical interpretation of this term is shown
schematically in fig \ref{area}.
\begin{figure}
\begin{center}
\epsfig{file=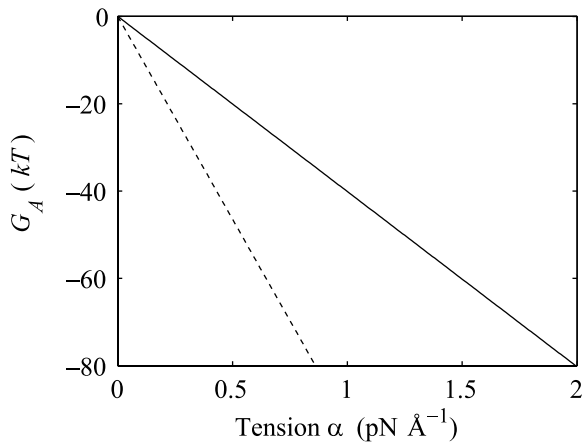}
\caption{\label{areade} 
The theoretical areal deformation free energies for the open (dashed line) and 
closed state (solid line) as a function of applied tension.}
\end{center}
\end{figure}
The form of this contribution is well understood (for example, see  Hamill and
Martinac, 2001) and  
 is analogous to the $-PdV$ term for an ideal gas in three dimensions. 
For areal deformations, the bilayer lipids act like a two-dimensional gas 
with a free energy change given by
\begin{equation}
dG_A = -\alpha\, dA
\end{equation}
where $\alpha$ is the tension. At high tension, the open state is favored due to its larger area. 
Sukharev {\it et al.} (1999) have measured the opening tension to be 
$\alpha_{*} = 1.2\ {\rm pN\  \AA}^{-1}$. ($\alpha_{*}$ is the 
tension at which the channel is open half the time. This tension will depend on the bilayer 
in which the
channel is reconstituted but we use this number as the typical size of the opening tension.) 
The areal deformation energy is 
\begin{equation}
G_A = -\alpha A = -\alpha\ \pi R^2 \approx  -47 \left(\frac{\alpha}{\alpha_*}\right)\left(\frac{A}{A_C}\right)\ kT,
\end{equation}
and is plotted as a function of applied tension in figure \ref{areade}. The way in which this free energy is expressed is to normalize 
the tension in units of the opening tension, $\alpha_*$,  and the area in terms of the closed state area, 
$A_C$.  As we expect, the typical free energies generated by radial changes are large. This is no surprise 
since  the tension acts as the switch between the closed state and the larger open state. 
The most striking feature of this energy in comparison with those we will discuss below is its areal 
dependence. This free energy scales as the square of the channel radius, whereas almost all other contributions 
will roughly scale as the circumference. This scaling difference has important consequences for the 
stability of the conductance states and will give rise to a picture of the 
tension-induced opening of the channel much like the picture used to discuss nucleation of second phases.
We have gone to some length to develop the importance of this
scaling difference in our previous paper (Wiggins and Phillips, 2004).

Experimental measurements have roughly confirmed the linear dependence of the free energy difference on tension 
(Sukharev {\it et al.}, 1999). This would suggest that the open and closed states are relatively well defined,
at least with respect to the channel radius. If the closed state, for example, actually consisted of a 
heterogeneous mix of states, this would lead the dependence of the free energy on tension to deviate from
the linear relation predicted above. The fact that this has not been seen, indicates that well defined states 
are compatible with experiment.

\subsection{ Gaussian Curvature}
Gaussian curvature normally contributes to the free energy topologically (independent of the local shape 
of the bilayer.) However, at the inclusion, the bilayer has a boundary which will allow non-topological 
contributions to the free energy (E. Evans, personal communication). In the small-angle limit,
the Gaussian curvature free energy
is 
\begin{equation}
G_G = -\pi K_G\left(U'^{\, 2}+H'^{\, 2}\right)
\end{equation} 
as demonstrated in the appendix. Measurements of the Gaussian curvature modulus are compatible with 
a wide range of values: 
$K_G<-K_B/2$ (see references in Boal, 2002). We estimate that for MscL, the free energy contribution from the 
mid-pane slope is 
\begin{equation}
G_G \approx 0.7\left(\frac{-K_G}{K_B}\right) \left(\frac{H'}{0.1}\right)^{2}\ kT,
\end{equation} 
where the deformation energy has been written in a dimensionless form in terms of the bending modulus, $K_B$, the 
closed state radius, and a modest interface angle of $0.1$. (We expect the contribution from $U'$ to be of the 
same order.) We have chosen this small angle since a large 
tilt angle for the interface is not evident from the closed state structure or the modeled open state
(see figure \ref{rendered}).
%which seems reasonable since the same physical material properties give rise to
%both moduli. 
As indicated above, the free energy is typically fairly small unless $H'$ or $U'$ are 
large. Since $G_G$ is radially independent, it induces no tension. On the other hand, Gaussian curvature does 
induce a torque of the form
\begin{equation}
\tau_\pm = \pi K_G H'_\pm,
\end{equation}
which points toward $H'_\pm=0$ if $K_G<0$. The induced mid-plane torque is exactly what is expected 
from the bending stresses in eqn. \ref{bs}.

\subsection{ Spontaneous Curvature}
\begin{figure} 
\begin{center}
\epsfig{file=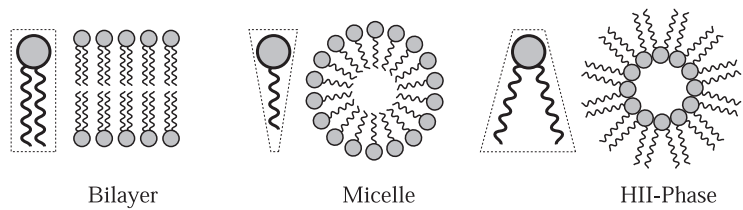}
\caption{ \label{lipids} A schematic depiction of molecular shapes which influence spontaneous curvature 
(Israelachvili, 1991).
Molecules with a cylindrical shape, such as phospholipids, will assemble into bilayers. 
Cone shaped molecules, such as lysophospholipids will assemble into micelles, the 
lowest energy configurations. For our sign conventions, these cone shaped molecules induce 
negative spontaneous curvature. Inverted cone shape molecules, such as cholesterol, DOPC, and DOPE
assemble into $H_{II}$ phases (Gruner, 1989) and induce positive spontaneous curvatures. 
The size of the spontaneous curvature is thought to be related to the  difference in size between
the polar head group and the acyl tails. Figure adapted from Lundb{\ae}k and Andersen, 1994.
}
\end{center}
\end{figure}

Spontaneous curvature arises from the addition of detergents and non-bilayer forming lipids 
to the bilayer. These molecular additions cause the lowest 
energy configuration of a single layer of lipids to be curved. The general phenomenon of spontaneous
curvature in lipid structures is reviewed by  Gruner (1989).
In general, measurements of the spontaneous curvature, $C$, have been for $H_{II}$ phase forming
molecules where the positive spontaneous curvature can be deduced from the lattice structure
(Gruner, 1989; Chen and Rand, 1997; Keller {\it et al.}, 1993). Values of $C^{-1}=20\ {\rm\AA}^{-1}$ 
(DOPE) are experimentally attainable (Keller {\it et al.} 1993). Less is known  about negative
spontaneous curvature, induced by micelle-inducing detergents and lysophospholipids. See fig. \ref{lipids}
for a brief discussion of molecular shape and spontaneous curvature.
To induce a composite bending modulus for the bilayer, the layers must be asymmetrically doped, though 
the molecules can exchange between the leaflets and move within a leaflet to energetically 
favorable locations caused by localized regions of high complementary curvature (de Kruijff {\it et al.}, 
1977; Kumar at al., 1989). For the sake of making an explicit estimate, we ignore 
these complications.
\begin{figure} 
\begin{center}
\epsfig{file=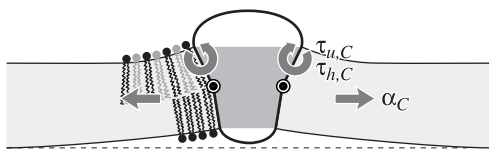}
\caption{ \label{spont} A schematic depiction of spontaneous curvature induced
by several species of lipids in the bilayer. The gray colored 
lipids depict a non-bilayer lipid which induces positive spontaneous curvature. A tilted inclusion 
interface can lead to a reduction in the stress caused by the non-bilayer lipids as depicted above.  
Spontaneous curvature induces both torques and tension at the interface. For energetically favorable
tilt, the tension acts to open the channel. The torque on the inclusion from a bilayer leaflet with positive 
spontaneous curvature acts to increase tilt by expansive pressure at the surface and compressive pressure
at the mid-plane. When only one leaflet of the bilayer is doped, both a mid-plane and a shape torque 
are induced but they cancel for the undoped leaflet. 
}
\end{center}
\end{figure}
\begin{figure} 
\begin{center}
\epsfig{file=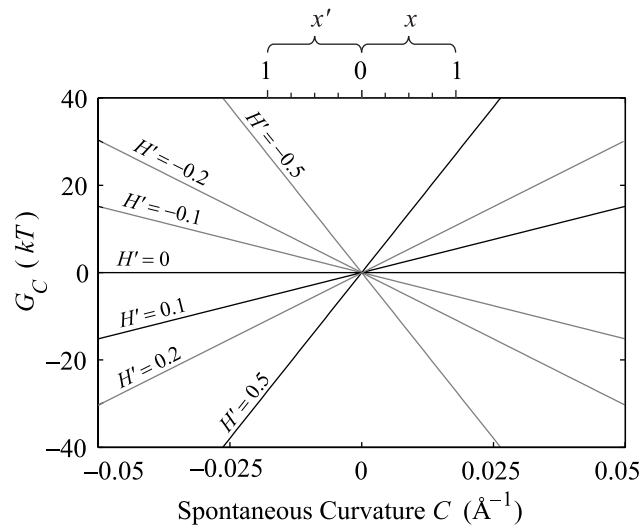}
\caption{\label{gc} 
The spontaneous curvature free energy as a function of the composite spontaneous curvature $C$ for various mid-plane slopes. 
At the top we have 
shown the corresponding concentration ratio for the DOPE/DOPC system of Keller {\it et al.} (1993).
For positive $C$, the bottom leaflet consists of pure DOPC and the top leaflet is a DOPE/DOPC mix with mole fraction $x$ of DOPE.
For negative $C$, the top leaflet consists of pure DOPC and the bottom leaflet is a DOPE/DOPC mix with mole fraction $x'$ of DOPE.
We have plotted the free energy for a range of spontaneous curvatures that are larger than those that can be realized for DOPC/DOPE
bilayers, since they may be relevant for other lipids or detergent-lipid bilayers. 
}
\end{center}
\end{figure}

In the linearized theory, the spontaneous curvature contributes
an interface term to the free energy. In fig. \ref{spont} an energetically favorable 
curvature is depicted. The free energy arising from spontaneous curvature is
\begin{equation}
G_{C} = 2\pi R K_B \left(CH'+\overline{C}U'\right) \approx 15 \left(\frac{R}{23\ {\rm \AA}}\right)
\left(\frac{20\ {\rm \AA}}{C^{-1}, {\overline{C}}^{-1}}\right) \left( \frac{H',U'}{0.1}\right)\ kT.
\end{equation}
These symbolic results are equivalent to those in 
Dan and Safran (1998). To estimate the size of this contribution for MscL, we have written the 
free energy in a dimensionless form using 
the large positive spontaneous curvature of a DOPE monolayer ($C^{-1}=20\ $\AA) (Keller 
{\it et al.} 1993), a relatively modest tilt angle ($H'=0.1$), and the closed state radius. The comma
notation is meant to denote that this estimate is for either these values of $C$ and $H'$ 
or $\overline{C}$ and $U'$. 
The resulting free energy 
can be the same order of magnitude as the areal deformation energy, implying it
may play an important role in channel function.  

Physically, the scaling can be easily understood with the example schematically 
illustrated in fig. \ref{spont}. A protein that has a conical shape, which increases 
toward the periplasm, induces membrane stress that may be relieved by complementary shaped lipids (which give 
rise to a positive composite spontaneous curvature) as illustrated in the figure.
The bilayer illustrated in the figure also has positive mean spontaneous curvature ($\overline C>0$), 
which relieves the stress induced by the hour glass shaped inclusion. 
This deformation energy is our first example of a line tension 
(a free energy with a linear radial dependence.)
This deformation energy is caused by interaction at the protein interface whose size is 
proportional to the interface area and therefore proportional to the radius of the inclusion. 
We have described in detail the significance of this linear dependence for mechanotransduction 
elsewhere (Wiggins and Phillips, 2004).

Spontaneous curvature gives rise to both a tension, due to the radial 
dependence of the free energy, and torques, due to the dependence of the free energy on $H'$ and $U'$. The 
tension on the boundary of the protein is
\begin{eqnarray}
\alpha_C = -\frac{K_B}{R} \left(CH'+\overline{C}U'\right) \approx -0.19 \left(\frac{23\ {\rm\AA}}{R}\right) 
\left(\frac{23\ {\rm\AA}}{C^{-1},{\overline{C}}^{-1}}\right)\left(\frac{H',U'}{0.1}\right),
\end{eqnarray}
where we have estimated the size of the induced tension by writing it in a dimensionless form using the same parameters  
as the deformation energy described above. This induced tension can have either sign resulting in contributions which are either tensile or compressive.
If curvature stress is relieved by spontaneous curvature, it is energetically 
favorable to increase the radius and
the tension tends to open the channel while if the curvature stress is increased by the spontaneous curvature,
the tension will be compressive. 
The mid-plane torque is
\begin{equation}
\tau_{h,C} = -2\pi R K_B C \approx  -1.5\times 10^2\ \left(\frac{R}{23\ 
{\rm \AA}}\right)\left(\frac{20\ {\rm \AA}}{C^{-1}}\right)\ kT,
\end{equation}
which is again written in a dimensionless form as described above. The torque induces inclusion 
conformations that would allow energetically favorable bending as explained above
and depicted in fig. \ref{spont}. The mid-plane torque is non-zero only for asymmetrically doped 
bilayers and its symbolic form matches the spontaneous curvature term deduced from bending stress in 
equation \ref{bs}.
The shape torque is 
\begin{equation}
\tau_{u,C} = -2\pi R K_B{\overline C} \approx  -1.5\times 10^2\ \left(\frac{R}{23\ 
{\rm \AA}}\right)\left(\frac{20\ {\rm \AA}}{{\overline C}^{-1}}\right)\ kT,
\end{equation}
which, for positive mean spontaneous curvature, acts to compress the mid-plane and expand  the
outer surface region of the inclusion. (Again, we have written the torque in a dimensionless form
as described above.)

%These issues have been considered by 
Keller at al. (1993) have studied the Alamethicin channel reconstituted into DOPC/DOPE bilayers.
This is a particularly beautiful system since the spontaneous curvature of the mixed bilayer 
interpolates linearly with the relative concentration of the components, allowing a continuous 
range of spontaneous curvatures.
The bilayers of Keller {\it et al.} are symmetric, implying that $C=0$.  In this case, the spontaneous curvature 
free energy is (Dan and Safran, 1998)
\begin{equation}
G_C = 2\pi R K_B \overline{C} U',
\end{equation}
which would predict free energy differences between states to be linear in $\overline C$ which
Keller {\it et al.} (1993) have shown experimentally. 

%Furthermore the equation above predicts an energy shift of the
%right  size for reasonable protein geometry parameters. 
%(PW, unpublished) 

\subsection{ Bilayer Interface Energy}
\label{IET}
The bilayer and protein are glued together by hydrophobic-hydrophilic interaction 
forces which are strong enough to hold the protein in the bilayer at a typical cytoplasmic pressure
of several atmospheres. It is natural to assume that
in addition to the internal protein and bulk bilayer energies there will be an interaction term from the interface.
There are many complicated scenarios which might be dreamed up, but the simplest is to assume that there is
free energy proportional to the area of protein and bilayer in contact, resulting in a free energy
\begin{equation}
G_W = \sigma 2\pi R W
\end{equation}
where $W$ is the thickness of the hydrophobic region. The constant of proportionality, 
$\sigma$, is the interface energy and has units of energy/area. Thus far, we have concentrated 
exclusively on the bilayer bulk for two reasons:  (i) the continuum model is almost certainly a 
reasonable rough model for the processes of interest, and (ii) the material parameters for the bilayer
are known from earlier experiments (Rawicz {\it et al.} 2000). In contrast, little is known about the validity 
of this model for the interface nor is there any estimate for the size of $\sigma$, the interface energy. 
This class of interface terms gives rise to a
tension and a compressive-expansive force:
\begin{eqnarray}
\alpha_W &=& -\sigma \frac{W}{R}, \\
{F}_{W} &=& -\sigma 2\pi R. 
\end{eqnarray}
The effects of the tension and compressive force depend on the sign $\sigma$, the interface energy density. 
When $\sigma$ is positive, the interface is minimized, leading to compressive forces. When $\sigma$ is 
negative (the affinity of lipid and protein are high), the interface is maximized and the forces
are expansive.

We have introduced this energy as a sanity check for our boundary 
conditions. We have somewhat naively assumed that the membrane adapts to
an arbitrary protein shape. This assumption certainly fails when the 
adhesive forces attaching the membrane to the protein are not large enough 
to sustain the strain in the membrane. It is therefore useful to develop
an approximate expression for these adhesive forces.
We know the interface energy for a typical hydrophobic-hydrophilic mismatch 
(Hamill and Martinac, 2001)
\begin{equation}
\sigma_* = 25\, {\rm cal}\ {\rm mol}^{-1}\, {\rm \AA}^{-2} = 0.0418\, k_BT\,{\rm \AA}^{-2},
\end{equation}
which is large compared to the other tensions in the problem. 
The compressive force countering the creation of this interface is
\begin{equation}
{F}_{W*} = -\sigma_* 2\pi R \approx -2.5\times10^2\,\left(\frac{R}{23 \rm \AA}\right)\,{\rm pN}
\end{equation}
where we have used the closed state radius to write the force in a dimensionless form.
This force can be interpreted as the critical force required to strip the protein from the bilayer. 
As we have reasoned above, this force will be important when we consider the 
large deformation limit on forces and energies due to thickness deformation.

\subsection{ Mid-Plane Deformation }
\begin{figure} 
\begin{center}
\epsfig{file=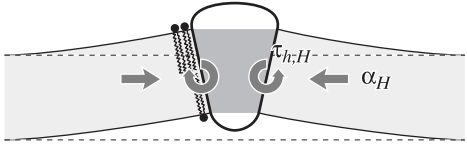}
\caption{\label{bend} A conically shaped protein induces bilayer bending. In order to 
match a conical inclusion interface, the bilayer must deform. The deformation leads to 
energetic contributions both from an increase in bilayer area and from bilayer bending. 
Mid-plane deformation induces both a mid-plane torque and a tension. The tension is 
always compressive. The mid-plane torque acts to reduce interface tilt and restore the
bilayer to its undeformed configuration.  We estimate that the mid-plane deformation
energy is probably not important for MscL gating. }
\end{center}
\end{figure}

The free energy associated with the deformation of the mid-plane of the bilayer
is another contribution in the overall free energy budget. 
These constant thickness deformations like
those pictured in fig. \ref{bend}, are induced by conically shaped proteins.
Mid-plane deformation contributes to the bilayer deformation energy  
through both bending of the bilayer and from a corresponding 
increase in bilayer area.
\begin{figure} 
\begin{center}
\epsfig{file=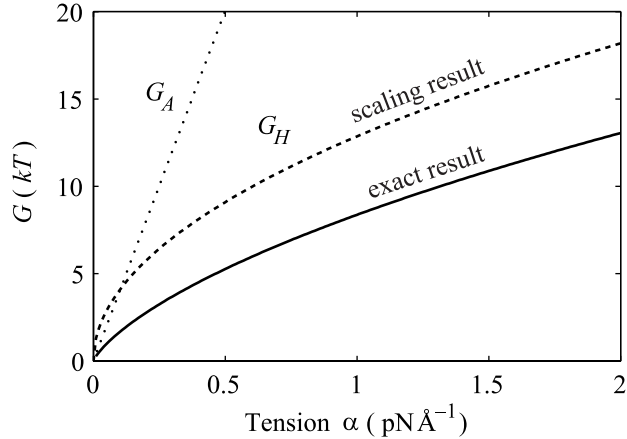}
\caption{\label{gh} 
The mid-plane deformation energy is illustrated above as a function of tension. We have plotted 
the approximate scaling result (dashed line) discussed below, the exact result to the 
model (solid line) discusses in the appendix \ref{gha}, as well as the areal deformation energy 
for the closed state, the opening tension $\alpha^*$
(dotted line). All the energies are computed for the closed state using an unrealistically large 
mid-plane slope (H=0.5)  to exaggerate the effect. While the scaling result is several
 $kT$ larger than the exact result, it accurately reflects the scaling at high tension, and 
provides a limit for the exact result. The $\alpha^{1/2}$ dependence 
of the mid-plane deformation energy has not been observed experimentally.   
}
\end{center}
\end{figure}
The exact result to the linearized model is derived in the appendix, 
but the dominant contribution at high applied 
tension is given by
\begin{equation}
G_{H} = \pi R \sqrt{\alpha K_B} \left(H'\right)^2 \approx 0.6 \left(\frac{R}{23 \AA}\right) 
\left(\frac{2a}{40.7 \AA}\right)^{3/2} \left(\frac{\alpha}{\alpha^*}\right)^{1/2}
\left(\frac{H'}{0.1}\right)^2 kT,
\end{equation}
where the parameters used to write the deformation energy in a dimensionless form are the closed state radius, the 
opening tension, and a modest interface tilt angle ($H'=0.1$).
The $H'^2$ dependence of the mid-plane deformation energy is as one would expect since no bending corresponds 
to $H'=0$ and results in the minimum energy (in the absence of spontaneous curvature). Dan and Safran (1998) 
have discussed deformation energies with a similar 
dependence on $H'$, but with a different size and physical origin. 
Note that mid-plane deformation scales differently with the
applied tension ($\alpha^{1/2}$) from the other contributions and can therefore 
be distinguished from the other bilayer deformation energies by measuring the tension dependence of the free energy.
%To estimate the size of this bilayer deformation energy, we have used $\alpha=\alpha_*$ and used the radius
%of the channel in the closed state when computing the area factor.
The approximation we have used is not really valid for MscL at experimentally realizable tensions since the elastic
decay length is given by 
\begin{equation} 
\sqrt{\frac{K_B}{\alpha}} \approx 27\left(\frac{2a}{40.7\,\rm\AA}\right)^{3/2}
\left(\frac{\alpha_*}{\alpha}\right)^{1/2}\ 
\rm \AA,
\end{equation}
where we have estimated the typical size of the decay length by writing it in a dimensionless form using the opening tension.
(The size and scaling of the bending moduli are described in the appendix.)
This decay length is roughly the same size as the channel radius. At high tension this length scale is
reduced thus improving the asymptotic result and also increasing the size of the energy. For MscL, unless 
the bending modulus is significantly softened, we are unlikely to be able access this regime since 
the lysis tension for bilayers is typically not much more than $\alpha_*$ (Olbrich {\it et al.}, 2000).  
The scaling result we have derived overestimates the bilayer deformation energy. 
(See the appendix for further discussion.) Both the exact result and asymptotic result are plotted as a function
of applied tension in figure \ref{gh}.
In spite of this overestimate, the energy
is still small compared with the areal deformation, so we conclude 
that mid-plane deformation is probably not a key player in the free energy budget for MscL.  
This effect has also been explored in a recent paper by Turner and Sens (2004). 

%The dominant term in mid-plane deformation 
%energy scales linearly with $R$ since the area of 
%the bilayer deformation is roughly proportional to the circumference. Note that this 
%implies that even if $H'$ remains constant, but nonzero, 
%in a transition, the free energy can still change if the radius changes.  Further,
%if the midplane deformation changes during channel gating, for example as a result of
%the tilting of transmembrane alpha helices, this can induce a larger contribution to
%the free energy.  This effect has been explored in a recent paper (Turner and Sens, 2004).

The dominant term in the mid-plane deformation energy scales linearly with $R$ since the area of 
the bilayer deformation is roughly proportional to the circumference.
This radial dependence gives rise to a tension:
\begin{equation}
\alpha_H  = \frac{1}{2R} \sqrt{K_B \alpha} \, H'^2 \approx 7.0\times 10^{-3} \left(\frac{23\, \rm \AA}{R}\right) 
\sqrt{\frac{\alpha}{\alpha^*}} 
\left(\frac{H'}{0.1}\right)^2\, {\rm pN}\, {\rm \AA}^{-1},
\end{equation}
which acts to inhibit channel opening. For the typical constants chosen here, 
$\alpha_H$ is about a hundredth of the opening tension, again confirming that the mid-plane deformation is
probably not important for MscL conformation or function.
Like the spontaneous and Gaussian curvature contributions, the mid-plane deformation also 
places a torque on the protein: 
\begin{equation}
\tau_H = -2\pi R \sqrt{\alpha K_B} H' \approx -11 \left(\frac{R}{23 \AA}\right) \sqrt{\frac{\alpha}{\alpha^*}} 
\left(\frac{H'}{0.1}\right)\ kT,
\end{equation}
which we have written in a dimensionless form as described above. 
This is a restoring torque toward the lowest energy configuration $H'=0$ (in the absence of spontaneous 
curvature.)

\subsection{ Thickness Deformation}
\label{TDT}
\begin{figure}
\begin{center}
\epsfig{file=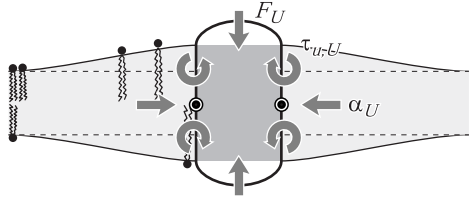}
\caption{ \label{thick} Bilayer thickness deformation due to a hydrophobic mismatch. 
In order to match the inclusion's hydrophobic boundary, the bilayer thickness must be 
deformed. Microscopically, the lipid tails are deformed as illustrated schematically 
above. The modulus for these deformations is $K_A$. For large mismatches, the energy 
contribution from thickness deformation can be quite significant. We estimate that this 
energy is important for MscL gating. Thickness deformation induces a 
compression-expansion force, a tension, and a shape torque which are also depicted above. 
The compression-expansion force acts to reduce the mismatch. The shape torque acts to induce interface tilt to
reduce the bilayer bending. The tension generated by a mismatch is always compressive. } 
\end{center}
\end{figure}
%\begin{figure}
%\begin{center}
%\epsfig{file=figures/rendered.eps}
%\caption{ \label{rendered} Above are the closed and open state models constructed by 
%Sukharev {\it et al.} (200?). 
%The grid in the figure has $10\ \rm \AA$ spacing. We 
%estimate from these structures that the hydrophobic thickness of the closed state is 
%roughly $W=38\ \rm \AA$ maximum protein-lipid coupling as measured by Powl {\it et al.} (200?).... } 
%\end{center}
%\end{figure}
The free energy contribution from thickness deformation results from changes
in  the separation between the upper and lower surfaces of the bilayer induced by the 
hydrophobic mismatch between the inclusion and the bilayer. 
This effect is depicted schematically in fig. \ref{thick}. 
The energetic contribution from this mismatch can be quite significant. 
Bilayer thickness deformation 
has been studied by many authors: Mouritsen and Bloom (1984), Huang (1986), etc. and more 
recently by Nielsen {\it et al.} (1998) and Goulian {\it et al.} (1998). These authors have all 
solved the model exactly, but 
we introduce a large radius asymptotic expansion to significantly simplify our results.
Expanding the exact solution of the model in powers of the radius gives
\begin{equation}
G_U = G_U^{(0)}+G_U^{(1)}+...
\end{equation} 
where $G_U^{(n)} \propto R^{1-n}$. For MscL, the only important terms are the first two. 
In the appendix we have plotted both the approximate and exact solutions to demonstrate 
that the interesting physics is captured by our approximations.   
Ignoring higher order terms, the resulting contribution is 
\begin{equation}
G_{U} = \pi R \left[ K_B\left(\beta_++\beta_-\right)\left(U'+ \left[\beta_++{\textstyle\frac{1}{2R}}\right] 
U\right)\left(U' + \left[\beta_-+{\textstyle\frac{1}{2R}}\right] U\right)-\alpha UU' \right], 
\label{exactthickness2}
\end{equation}
where
\begin{equation}
\beta_{\pm} \equiv \sqrt{ \frac{\alpha\pm\sqrt{\alpha^2-4K_B K_A/a^2}}{2K_B}}.
\end{equation}
We can simplify this expression further by defining a low tension limit (Goulian {\it et al.}, 1998):
\begin{equation}
\alpha \ll 2\sqrt{\frac{K_B K_A}{a^2}} \\ \approx 
0.34 \left(\frac{2a}{40.7\,\rm\AA}\right)\, kT\,{\rm \AA}^{-2} \sim 10\alpha_*
\end{equation}
which is roughly satisfied for 
the critical tension measured by Sukharev {\it et al.} (1999). 
(See the appendix for details about the scaling and size of the elastic moduli. The tension
above has been put into a dimensionless form using the parameters for a PC lipid of acyl length 18.) 
One might worry that for small bilayer thickness the small tension  limit would no longer
be satisfied, but we will show that the opening tension 
is also reduced in this case. In the
low tension limit, $\beta_\pm$ is
\begin{equation}
\beta_{\pm} = e^{\pm \frac{i\pi}{4}} \left(\frac{K_A}{a^2 K_B}\right)^{\frac{1}{4}} = e^{\pm \frac{i\pi}{4}} \beta.
\end{equation} 
We will refer to $\beta$ as the inverse decay length since it defines the length scale over which the 
thickness deformation perturbation decays.  This length scale is given by
\begin{equation}
\beta^{-1} = \left(\frac{K_Ba^2}{K_A}\right)^{1/4} \approx 11\left(\frac{2a}{40.7\, \rm\AA}\right)\ {\rm \AA}.
\end{equation}
This decay length defines the large radius limit, which is satisfied even for the closed state of MscL since
$R_C>\beta^{-1}$. 

The dominant contribution at large radius is $G_U^{(0)}$, which corresponds to ignoring the curvature of the 
interface entirely (Dan {\it et al.}, 1993).  To estimate the typical size of this contribution, we set $U'=0$:  
\begin{equation}
G_U^{(0)} = \pi R {\cal K}U^2 \approx 1.6 \left(\frac{R}{23 \,\rm \AA}\right) \left(\frac{U}{1 \,\rm \AA}\right)^2\
kT,
\end{equation}
where we have written the deformation energy in a dimensionless form using the closed state radius, a small mismatch ($U=1\ \rm \AA$), 
and the effective elastic modulus ${\cal K}$, defined
\begin{equation}
 {\cal K} \equiv \sqrt{2}\left(\frac{K_B K_A^3}{a^6}\right)^{\frac{1}{4}} \approx 2.16 \times 10^{-2}\,kT\,{\rm\AA}^{-3}.
\end{equation} 
This is the result listed in table \ref{table}. 
Since large mismatches are possible and the deformation energy grows as the square of the mismatch, 
this contribution can be very significant. This $U^2$ dependence, analogous to a linear spring,  
is exactly what we expect since the minimum energy occurs for a perfect thickness match between 
the protein and bilayer ($U=0$). Mouritsen and Bloom (1984, 1993) were the first to discuss this 
dependence. Its phenomenological significance has been stressed by Lundb{\ae}k {\it et al.} (1996).  
The thickness deformation energy is a function of both $K_A$, the local thickness deformation modulus, 
and the bending modulus, $K_B$.  Physically, $K_B$ provides a compatibility condition for 
adjacent lipids which sets the size of the deformed region. The thickness deformation
energy  is also roughly linear in $R$ since the area of the bilayer deformed is roughly 
proportional to the channel circumference. The size of thickness deformation energy and its radial 
dependence imply that $G_U^0$ is almost certainly important in the energetics of MscL. 
Unless both $U'$ and $U$ are zero or cancel, this term will contribute due to the radial expansion
of the channel between the closed and open states.
Even if the height of the hydrophobic region were to remain unchanged, this contribution
would still be very significant. (See Wiggins and Phillips, 2004) 
Let us mention, as a brief aside, that the functional form of $G_U^0$ is a
very pleasing result since, although the prefactor $\cal K$ appears to depend on the bilayer width, 
$a$, it is roughly independent of $a$! Please see the appendix for a brief argument. Because this
scaling is not obvious and we will often use scaling arguments, we write the result in terms of 
${\cal K}$ to alleviate the temptation of thinking ${\cal K} \propto a^{-3/2}$.

One of the difficulties in implementing this model is the uncertainty concerning
boundary conditions and in particular what slopes should
be assigned for the bilayer-inclusion interface.
One of the possibilities studied by other authors (Helfrich and Jakobsson, 1990),
is to treat $U'$ as a free parameter and minimize the free energy with respect to it.
In the asymptotic limit this calculation becomes very simple.
Taking the low tension limit ($\alpha=0$), and choosing $U'$ to minimize $G_U^{(0)}$, 
gives a free energy half that which is obtained by naively assuming $U'=0$, namely,  
\begin{equation}
G_U^{(0),\rm Min} = \frac{\pi R}{2} {\cal K} U^2.
\end{equation}
As a result, we argue that the qualitative conclusions--the importance of 
this correction--are indifferent to the particular
choice made for this boundary condition, but quantitatively the choice of 
boundary conditions can have a significant effect.

Although $G_U^{(0)}$ dominates at large radius, for MscL-like geometries, 
$G_U^{(1)}$ is roughly as large. $G_U^{(1)}$, which is radially independent, is
\begin{equation}
G^{(1)}_U = 2\pi\left(\frac{K_B K_A}{a^2}\right)^{1/2}U^2 \approx 1.1\ \left(\frac{2a}{40.7\ {\rm \AA}} 
\right) \left(\frac{U}{1\ {\rm \AA}}\right)^2\ kT,
\end{equation}   
for $U'=0$, where the energy has been put in a dimensionless form using the closed state radius, and a small
mismatch ($U=1\ \rm \AA$). 
As can be seen above, for the closed state, this energy is almost as large as the dominant scaling 
term $G_U^{(0)}$ and is also proportional to $U^2$. In general, the effects of this term on channel 
gating are not as pronounced since it is radially independent and will not contribute a term to the 
free energy difference between the open and closed states proportional to $\Delta R$. Likewise, it 
will not contribute to the tension. The asymptotic expressions for the thickness deformation 
energy are compared with the exact results to the linearized theory in figure \ref{thickness} 
in the appendix. We plot the thickness deformation energy for the closed state in figure 
\ref{saturation_data}, in the next section.

To develop physical intuition into how thickness deformation affects the channel conformation and function, 
we calculate the generalized forces induced on the inclusion. The tension is:
\begin{equation}
\alpha_U = -\frac{{\cal K}U^2}{2R} \approx -2.0\times 10^{-2} \left(\frac{23\, \rm \AA}{R}\right)\left(\frac{U}{1 
\,\rm \AA}\right)^2\, \frac{\rm pN}{{\rm \AA}},
\end{equation}
which has been written in a dimensionless form as described above. The induced tension
acts to close the channel. 
For a $1 \rm \AA$ mismatch, the tension is roughly a sixtieth of 
what Sukharev {\it et al.} (1999) measured for the opening tension, but for larger mismatches, the tensions can become 
comparable, significantly reducing the net tension or, at small enough applied tension, becoming the 
dominant contribution. Since tensions of this size are responsible for triggering the channel to 
switch from the closed to the open conformation, this calculation strongly suggests that the 
thickness deformation energy plays an important role in channel function and conformation.   
The thickness deformation also generates a shape torque:
\begin{equation}
\tau_{u,U} = -2\pi R \frac{\sqrt{K_A K_B}}{a}\left(1+\frac{1}{\sqrt{2}\beta R} \right) U 
\end{equation}
when $U'=0$.  We can estimate the dominant term at large radius
\begin{equation}
\tau_{u,U} \approx -25\left(\frac{R}{23\ {\rm \AA}}\right) \left(\frac{2a}{40.7\ {\rm \AA}}\right)\left(\frac{U}{1\ {\rm \AA}}\right)\ kT,
\end{equation}
which has been written in a dimensionless form as described above.
The shape torque can be quite large for large mismatches. Its sign depends on the mismatch $U$.

\subsection{ Saturation of Thickness Deformation }
\begin{figure}
\begin{center}
\epsfig{file=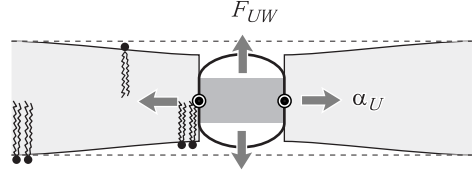}
\caption{ \label{saturation} Bilayer thickness deformation saturates when the energy required to
further deform the membrane is equal to the interface energy required to create a 
hydrophobic-hydrophilic interface. This failure of the bilayer to conform to the protein
is depicted schematically above. }
\end{center}
\end{figure}
\begin{figure}
\begin{center}
\epsfig{file=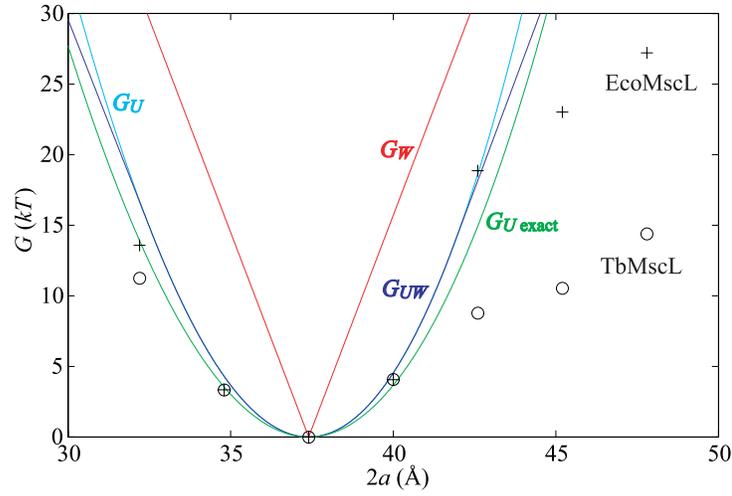}
\caption{ 
\label{saturation_data} 
Interface and thickness deformation energy of the closed state compared to experimental data
from Powl {\it et al.} (2003) as a function of lipid bilayer thickness. The red curve is the hydrophobic 
interface energy ($G_W$) without thickness deformation (the limit as $\cal K\rightarrow \infty$.) The green curve
is the exact thickness deformation energy ($G_{U\ \rm exact}$) without saturation. The cyan curve is the 
asymptotic thickness deformation energy  ($G_U$) without saturation (the limit as $\sigma_*\rightarrow\infty$.) 
The blue curve is the saturating thickness deformation energy  ($G_{UW}$, see the appendix for details). 
The $o$'s and $+$'s are the experimental
values measured by Powl and coworkers for TbMscL and EcoMscL respectively. 
We have chosen the closed state thickness of the channel ($W_C=37.5$ \AA) to match the thickness of the 
bilayer at the minimum of the experimental bilayer deformation energy. This thickness is 
compatible with the known closed state structure.
For small mismatches there 
is a much better qualitative agreement between the thickness deformation energy than the hydrophobic 
interface theory. For large mismatch, the experimental data points are significantly smaller than the 
energy predicted by theory. We discuss this apparent discrepancy in the next section. 
}
\end{center}
\end{figure}
Due to the quadratic dependence of the thickness deformation energy on mismatch, it is initially energetically 
favorable to deform the thickness of bilayer, rather than expose the hydrophobic region of the protein to the solvent. 
But, this quadratic dependence also implies that the energetic cost of further deformation will continue to grow, 
until, at a critical mismatch, it becomes more costly than exposing this added region to the solvent. 
This critical mismatch is related to the compression force on the inclusion due to the thickness deformation. 
Recalling that $U\equiv W/2-a$, the compressive force on the protein is
\begin{equation}
F_U = -\frac{\pi R}{2} {\cal K}\left(W-2a\right)\left(1+{\textstyle \frac{\sqrt{2}}{\beta R}}\right)
\approx -54\left(\frac{R}{23\rm \AA}\right)\left(\frac{W-2a}{1 \rm \AA}\right)\ {\rm pN},
\end{equation}
which has been written in a dimensionless form as described above.
The change in the thickness deformation energy for increasing the hydrophobic region of the protein from 
$W$ to 
$W+dW$ is $-F_UdW$ whereas to expose the added region to solvent results in an energy increase of $-F_{W*}dW$.
At the critical mismatch, these two forces are equal:
\begin{equation}
F_{W*} = F_U(W).
\end{equation}
Solving for $2U$ gives the critical mismatch:
\begin{equation}
2U_* \equiv \left|W-2a\right| = \frac{4\sigma_*}{{\cal K}{\left(1+\frac{\sqrt{2}}{\beta R}\right)}} \approx 5\,\rm \AA,
\end{equation}
which has been estimated for an acyl length 18 PC lipid bilayer and the closed state radius.
The details of the saturated thickness deformation energy are worked out in
the appendix. This saturated deformation energy is compared to the thickness deformation energy and experimental 
deformation energies measured by Powl {\it et al.} (2003) in figure \ref{saturation_data}.
For large mismatch, there are discrepancies between the experimental data and all the theoretical models. It is 
unclear whether the lipid finds a more energetically efficient method for offsetting the mismatch. 
In principle lipid packing 
calculations could answer these types of questions, but typically they are too constrained to capture this type of 
behavior. We shall return to this question in the next section.
%of Fattal and Ben Shaull (1997?) have not predicted this effect, perhaps because their calculation 
%are too constrained.   

% does deviate from the curve 
%predicted by thickness deformation at roughly this mismatch, but as depicted in figure \ref{saturation_data}  
%the free 
%energy is much smaller than predicted by exposing the lipid to the solvent. 
%For large mismatch there is a discrepancy between the experimental data of Powl {\it et al.} (2003) and theory.
%It is unclear from this data whether the reason for this discrepancy is a conformational change in the protein
%or a saturation of the thickness deformation sooner than we have predicted.

Over the course of this entire section, we have provided a term-by-term dissection of the various
contributions to the free energy of deformation associated with channel gating.  The main point
of this exercise has been to provide a framework for thinking about the connection between
ion channel gating and the corresponding perturbations induced in the surrounding lipid
bilayer membrane.  

%As shown in several places during the discussion, our model provides
%a framework for interpreting experiments.    

\section{Application to MscL Gating}
\label{ATMG}
The conformational landscape of the MscL protein is certainly extremely complex, depending on a 
large number of microscopic 
degrees of freedom which are analytically intractable. 
Even from the standpoint of numerical calculations, this number
is still very large (Gullingsrud {\it et al.}, 2001). 
What is the point of examining what is presumably only half the story by treating the bilayer 
analytically? The purpose of this model is to pose a theoretical problem simple enough to be completely
soluble, yet not so simple that it bears too little resemblance to the complex system it represents. 
By understanding the consequences of the simplest models, we develop a framework in which to understand 
the richer dynamics of the real system, whether approximated by molecular 
dynamics simulations or studied in experiments. There is a wealth of useful, physical intuition to 
be gleaned from this model relating to both the function of the mechanosensitive channel (MscL) and 
that of mechanosensitive transmembrane proteins in general.

As we have argued in the previous section, the bilayer deformation energy is comparable to the measured
free energy differences between states for the MscL channel. Therefore the bilayer must play an important role 
in determining the free energy balance between states, altering the channel function.    
It is also likely that the forces generated by bilayer deformation can 
significantly perturb the conformation of the states themselves.  Indeed, to the
extent that membrane deformations induce conformational changes in the protein itself,
the structure of the protein itself becomes lipid context dependent, complicating predictions.
At present, we treat the protein as a black box which gives us a fixed geometry for state $i$ described 
by the state vector $X_i$, and a protein conformational free energy, $G_{P,i}$. As we have discussed above,
the geometry of the channel in the $i$th conformational state is described by the radius ($R_i$), the hydrophobic 
thickness ($W_i$), and the two
angular parameters that we usually interpret as the shape of the protein's interface ($U'_i,H'_i$). 
Please see fig. \ref{basic} and section \ref{nia}. 
These protein parameters are combined, for economy of notation, into the state vector $X_i$,  
\begin{equation}
X_i \equiv (R_i,W_i,U'_i,H'_i).
\end{equation}
We assume these
protein parameters are fixed by internal conformation and do not depend on the parameters of the bilayer
membrane such as the lengths of the lipid tails or the concentration
of spontaneous curvature inducing lipids, nor on the applied tension, $\alpha$. We will call this simplified picture the static conformation approximation.
Explicitly, we assume the free energy takes the form:
\begin{equation}
G_{i} = G_{P,i}+G_{\cal M}(X_i), 
\end{equation}
for state $i$ where $G_{P,i}$ and $X_i$ are independent of the bilayer parameters and the applied tension. 
%$G_{\cal M}$ is bilayer deformation free energy we have calculated.
In principle, we can try to determine the unknown state vectors, $X_i$,  by varying the membrane parameters
and the applied tension. 
Of course the primary advantage of the static conformation approximation is that it allows simple 
predictions to be made relating to the channel gating. This model is probably reasonable for 
relatively modest changes to the bilayer parameters provided that the free energy wells corresponding 
to the conductance states are relatively sharp and well defined with respect to changes in the state 
vector $X_i$.

\subsection{Opening Probabilities for Two State System}
The difference in
free energy between the open and closed states is defined as
\begin{equation}
\Delta G = -kT \log \frac{{\cal P}_O}{{\cal P}_C} = \Delta G_{\rm P}+\Delta G_{\cal M},
\end{equation}
where ${\cal P}_i$ is the probability of state $i$, and $\Delta$ here is the difference between 
open and closed.
Notice that this expression is independent of the free energies 
of the other states as a result of working with the ratio of the open and closed
probabilities. For ease of interpretation, it is convenient
to further subdivide the free energy by subtracting off the areal deformation
contribution such that 
\begin{equation}
\Delta G = \Delta G_{\rm P}+\Delta G_{\cal M}^0-\alpha \Delta A,
\end{equation}
where the $\Delta G_{\cal M}^0$ is the bilayer deformation energy less the areal contribution. 
Since we expect the only tension dependence to come from the linear areal
deformation term, the measured $\Delta G$ should be linear in tension. 
What would a non-linear behavior tell us? It would signal that there is additional tension
dependence in the terms above. Provided that we are convinced the bilayer terms are correct, 
it would signal that the static conformation approximation is failing: the conformation of the
state is tension dependent! Data from Sukharev {\it et al.} (1999) have shown that $\Delta G$
is at least fairly linear in tension. Assuming that linear dependence discussed 
above is correct, the slope with respect to 
tension of the free energy gives us the area change:
\begin{equation}
\label{deltaA}
\Delta A = -\frac{\partial \Delta G}{\partial \alpha} 
\end{equation}
and the free energy can be written in a convenient form (Hamill and Martinac, 2001)
\begin{equation}
\label{openingtension}
\Delta G = \Delta A\left(\alpha_{1/2}-\alpha\right),
\end{equation}
where $\alpha_{1/2}$ is the opening tension (where the probability of being open or closed is equal) and
is given by
\begin{equation}
\alpha_{1/2} = \frac{\Delta G_{\rm P}+\Delta G_{\cal M}'}{\Delta A} = \frac{\Delta G_0}{\Delta A},
\end{equation} 
where $\Delta G_0$ is the free energy change with the areal deformation 
contribution removed or alternatively the free energy difference at zero tension. 
When the free energy is written in terms of the opening tension (eqn \ref{openingtension}), it is clear that changes 
in the lipid parameters, such as the equilibrium thickness for example, lead to a simple
offset of the opening tension, leaving the dependence of the ratio of open to closed  probabilities versus applied tension otherwise 
unchanged, as Perozo {\it et al.} (2002a) have observed. This observed offset behavior 
is indirect evidence that the change in the area between the closed and open states is roughly independent 
of the bilayer parameters, implying that the open and closed states are fairly well defined, at least radially.
In the rest of the paper
we will refer to $\Delta G_0$ as the free energy difference, dispensing with the subscript.

In patch clamp experiments, the tension is controlled indirectly via the pipette pressure. The pressure and
tension are related using Laplace's law:
\begin{equation}
P = \frac{2\alpha}{{\cal R}},
\end{equation}
where $\cal R$ is the radius of curvature of the membrane patch.
Typically it is assumed that this curvature is roughly constant during the experiment (e.g. 
Hamill and Martinac, 2001) which implies that opening pressure is proportional to the opening free energy:
\begin{equation}
\label{Peqn}
\Delta G_0 = P_{1/2} \left[\frac{\partial G}{\partial P}\right]_{P_{1/2}},
\end{equation}
where the derivative of $G$ is expected to be constant since it is $\Delta A \, {\cal R}/2$.

\subsection{Mismatch and Gating}
\begin{figure}
\begin{center}
\epsfig{file=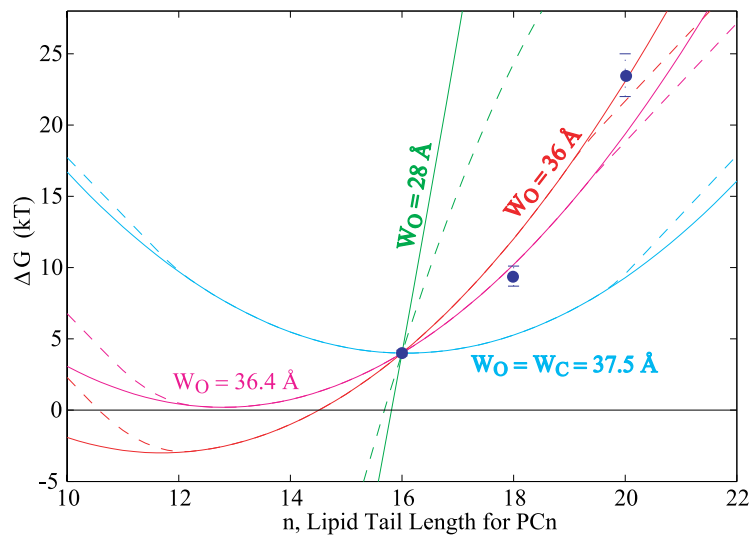}
\caption{ \label{perozodata} 
The theoretical free energy difference compared to the experimental data of Perozo {\it et al.} (2002a)
for different choices of the geometric parameters characterizing the open state thickness.
The thickness deformation energy is plotted for a closed state thickness of $W_C = 37.5\ \rm \AA$ and 
several different open state thicknesses. Each curve is shifted to pass through the data point at 
an acyl chain length of 16. An open state thickness of $W_O\sim 36\ \rm \AA$ give a reasonable
fit to the experimental data. Perozo and coworkers also have electron paramagnetic resonance data
for bilayers with acyl chain lengths $n\ge 10$ which suggest that the channel is closed ($\Delta G\ge 0$) in 
the absence of applied tension.  
}
\end{center}
\end{figure}

Before we begin our analysis in earnest, we wish to quickly remind the reader of the differences in the 
current model from that in our recent short paper (Wiggins and Phillips, 2004). In our previous paper we developed 
a simplified version of the model described above.  The only geometrical change between the open and closed 
states was in the channel radius. In that model, the energetics of the bilayer deformation energy is one dimensional
and can be analyzed as a competition between the bilayer line tension and the applied tension 
(Wiggins and Phillips, 2004):
\begin{equation}
G_{\cal M} = f\cdot 2\pi R-\alpha \cdot \pi R^2,
\end{equation}   
where $f$ is the line tension and where the only free parameter is the effective thickness of the 
protein which we fit using the data of Perozo {\it et al.} (2002a). On the other hand,
the simplifications associated with this model (i.e. we did not differentiate between 
the thickness of the open and closed states)  leave it unable 
to reproduce the data of Powl {\it et al.} (2003) which essentially measures the bilayer deformation of the 
closed state. In spite of this
limitation, this  simplest theory based upon the competition between the line tension and applied
tension reveals that (i) the acyl chain length dependence of the opening free energy as measured by 
Perozo {\it et al.} (2002a,b) is very naturally explained by the thickness deformation energy 
and can qualitatively explain that (ii) spontaneous curvature could lead to open state stabilization and that (iii) 
the substates of the 
channel should be short lived. In this section, we undertake a more quantitative analysis in which 
we allow the open and closed states to have different hydrophobic thicknesses.
In particular, we  analyze the experimental data from three different classes
of experiments in detail. First, we  focus on the 
opening free energy measurements by Perozo {\it et al.} (2002a). Next, we  analyze the lipid-MscL 
interaction data from Powl {\it et al.} (2003) and finally, we consider the recent mutation studies by 
Yoshimura {\it et al.} (2004) who altered amino acids in the transmembrane region of MscL.   Note
that we argue that our model should be viewed more as serving as an interpretive tool than
as a scheme for fitting experimental data.  As will be seen in the discussion to follow, the
act of interpreting the data from these various experiments consistently suggests that
the usual view of static protein structures that are lipid independent may have
to be amended. 

Perozo {\it et al.} (2002a) have measured the opening free energy of the channel 
for three bilayers with acyl chain lengths 
$16$, $18$, and $20$. 
We will fix the thickness of the closed state ($W_C=37.5\ \rm \AA$) based on experimental data from 
Powl {\it et al.} (2003) and corroborating computational evidence from Elmore and Dougherty (2003). 
This assignment seems reasonable based on the distribution of the hydrophobic residues in the 
closed state crystal structure as illustrated in  figure \ref{rendered}.
We now vary the open state thickness, $W_O$, and compare the resulting opening free energy versus lipid acyl 
chain length to the experimental data of Perozo {\it et al.} (2002a). While Perozo and coworkers have measured 
the opening free energy for only three acyl chain lengths, their electron paramagnetic resonance (EPR) 
data suggests that  even in acyl chain length 10 lipid bilayers, the channel does not open spontaneously in the absence 
of applied tension.
This qualitative information provides an additional constraint for the theory to satisfy ($\Delta G \ge 0$ for $n\ge 10$). 
We find that for $W_O \sim 36\ \rm \AA$, 
we have the best agreement with the experimental data. The comparison between the theoretical opening free energy 
and the measured values as a function of acyl chain length is depicted in figure \ref{perozodata}. 
Our fit with the experimental data is reasonable considering the complexity of the channel system and the naivete of
the static conformation model. The inability of the theory to fit the data exactly is to be expected from a model where 
the elastic constants have been fit to scaling laws and the subtle conformational changes of the protein are ignored.
As noted earlier, we view our model
as a framework for interpreting previous experiments and suggesting new ones, as well
as for providing intuition, rather than as a fitting scheme.
As is clear from the figure, it is quite difficult to satisfy both the large mismatch 
opening free energy for acyl length $n=20$ and the constraint that the channel be closed ($\Delta G \ge 0$) for 
acyl length $n \ge 10$. In light of the proposed structures for the open state (Sukharev {\it et al.}, 2001a,b; 
Betanzos {\it et al.}, 2001;
Perozo {\it et al.}, 2002a,b), our predicted change in channel thickness is quite modest. (See figure \ref{rendered}.) 
An open state with a smaller thickness satisfies neither the large $n$ nor the small $n$ limits.
 
We must treat the predictions of the theory with care when the mismatch is large since the theoretically predicted 
bilayer deformation energies are probably large enough to lead to protein conformational changes,
violating our static geometry approximation. That is to say, either or 
both the closed and open states of the protein deform significantly. This systematic uncertainty is not a peculiarity
of our models but a quite general uncertainty. For example, it is unclear that the lysophosphatidylcholine (LPC) 
stabilized conformation observed by Perozo {\it et al.} (2002a,b) is in precisely the same conformation as the 
open state of the channel stabilized by applied tension, reconstituted 
in a PC18 bilayer. One experimentally accessible probe to conformational changes is
a precise measurement of the applied tension dependence of the free energy difference between states. If the open and 
closed states are significantly perturbed by the applied tension, we would expect a deviation from the linear dependence 
(eqn \ref{deltaA}) of the free energy on applied tension. Alternatively, precise measurements of the area change between 
the open and closed states in different bilayers might show that the area change is lipid context dependent.  
We  revisit the question of conformational changes below.
At present, we conclude that experimental data of Perozo {\it et al.} (2002a,b) is compatible with the model.
Due to both the approximate nature of the static geometry approximation and the systematic uncertainties inherent in patch clamp measurements 
of channel opening free energies (E. Evans, personal communication) it is important not to place too high a premium on the precise fitting 
of the data of Perozo {\it et al.} (2002a).

\begin{figure}
\begin{center}
\epsfig{file=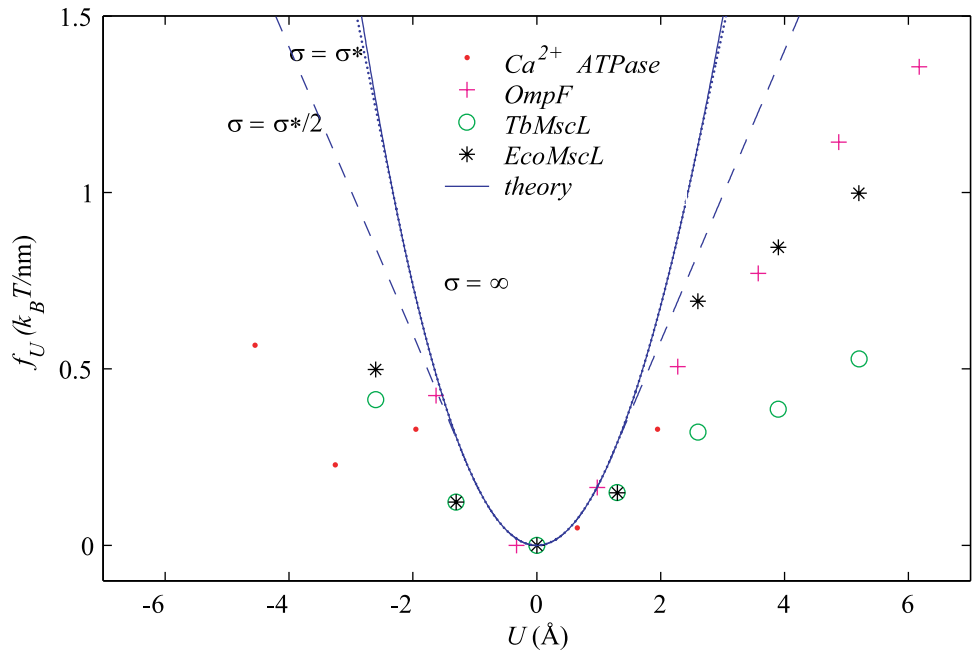}
\caption{ \label{sigma_n_powl} 
The theoretical line tension for MscL compared with the line tension estimated from the 
measurements of East, Lee and coworkers (East and Lee, 1982; O'Keeffe {\it et al.}, 2000; Powl {\it et al.}, 2003).
The experimental data for several different proteins has been aligned so that the minimum 
line tension is assumed to correspond to zero mismatch. In the small mismatch regime, there is very reasonable 
agreement between theory and the measurements. At large mismatch the story becomes more complicated. 
There is significant variation between proteins, and even between Eco and Tb MscL. These variations may signal 
conformational changes in the protein. 
The methods of East and Lee are only sensitive to the free energy in the first layer of lipids surrounding the proteins.
It is therefore natural to expect the theoretical line tension to be larger than the measured line tension. 
We have plotted the saturating thickness deformation energy ($G_{UW}$) for interface energies $\sigma =\infty$ (solid), $\sigma= \sigma_*$ 
(dotted), and $\sigma = \sigma_*/2$ (dashed) 
since $\sigma_*$ probably underestimates the saturation effect since the interface of the bilayer
which would initially be exposed to solvent is not extremely hydrophobic (e.g. White and Wimley, 1999).
}
\end{center}
\end{figure}
A more direct experimental method for analyzing the free energy of the MscL closed state has been exploited 
by Powl {\it et al.} (2003). East, Lee and coworkers (East and Lee, 1982; O'Keeffe {\it et al.}, 2000; Powl {\it et al.}, 2003)
have developed Trp fluorescence spectroscopy to study lipid-protein interactions. Their technique measures the lipid-protein 
binding constant for channels reconstituted in liposomes. The log of this binding constant is the free energy difference between 
lipids at the boundary and lipids
in the bulk of the bilayer (See Powl {\it et al.}, 2003 for details.) 
This free energy per lipid can then be converted into a line tension at the interface.
While this experimental technique provides a very direct measurement of the free energy per lipid, 
it is only sensitive to the free energy in the first layer of lipids surrounding the protein where there is direct interaction 
between lipid and protein.
Powl {\it et al.} (2003) measure a minimum line tension for an acyl chain length of 16, which roughly corresponds 
to a thickness 
of $37.5$ \AA. We assume that this chain length corresponds to zero mismatch, 
implying that the thickness of the closed state
equals the equilibrium thickness of the bilayer:
\begin{equation}
W_C = 2a_{16}. 
\end{equation}
We can now compute a theoretical line tension for the closed state as a function of acyl chain length.  
In figure \ref{sigma_n_powl}, we compare the experimental measurements of this line tension to the thickness deformation 
line tension predicted by theory.  In the small mismatch regime, there is very reasonable 
agreement between theory and measurement. This is a nontrivial result since although we have fit the data to choose
the minimum of the line tension, the curvature of the line tension (the second derivative with respect to the protein thickness)
is predicted by the bending moduli of the membrane measured at very small curvature on $\mu$m length scales!
At large mismatch, the predictions of the theoretical model are significantly higher than the experimentally measured values. 
There are several possible explanations for this discrepancy: (i) the predictions of the theory are too large for large mismatch
signaling the onset of nonlinear elastic effects, for example, 
(ii) the contribution from the energy not in the first ring of lipids surrounding the protein is large, 
or (iii) conformational changes in the protein reduce the size of the mismatch. 
For the moment, let us assume that the theory is incorrect for large mismatch (i). If we use the measured  
line tension, $f_U^{\rm \ exp}$, for a given mismatch, we can estimate the bilayer deformation free energy change 
between the open and closed states:
\begin{equation}
\Delta G' = [f^{\rm \ exp}_U \cdot 2\pi R]_O-[f^{\rm \ exp}_U \cdot 2\pi R]_C,
\end{equation}
where the $'$ is used to differentiate this computed free energy difference from that measured by Perozo {\it et al.} (2002a).
$\Delta G'$ can then be compared to the measured values of Perozo {\it et al.} (2002a) ($\Delta G$) with
the aim of examining the internal consistency of the model and both datasets.    For $W_O=36$ \AA, the free energies are
\begin{center}
\begin{tabular}{c|c|c|c|c}
$n$ & $\Delta G\ (kT)$ & $\Delta \Delta G\ (kT)$ & $\Delta G'\ (kT)$ & $\Delta \Delta G'\ (kT) $\\ 
\hline
$16$ &  $4$    & $0$    & $1.5$ & $0$ \\
$18$ &  $9.4$  & $5.4$  & $6.6$ & $5.1$ \\
$20$ &  $23.5$ & $19.5$ & $7.5$ & $6$
\end{tabular}
\end{center}
where $n$ is the acyl chain length, $\Delta G$ are the numbers measured by Perozo {\it et al.} (2002a) and $\Delta G'$ are those predicted
using the data of Powl {\it et al.} (2003). (See the appendix for more details on this calculation.) 
Remember that the numbers from Perozo are the total free energy change between states, 
the sum of both the membrane and protein free energy changes, while those we have estimated from the data of Powl include only 
the membrane interaction term. As before we will assume that the conformation and energy of the protein are roughly static, 
independent of the bilayer lipid acyl chain length. We therefore expect the free energy differences of Perozo and Powl to 
differ by a constant, corresponding to the protein conformational free energy difference, $\Delta G_P$. To eliminate the $\Delta G_P$ 
contribution, we examine the relative changes in the opening free energy relative to the opening free energy for 
the acyl chain length 16 bilayer: 
\begin{equation}
\Delta \Delta G \equiv \Delta G-\Delta G_{16}.
\end{equation}
The data of Powl {\it et al.} (2003) predicts the difference between
the acyl chain lengths $16$ and $18$ ($\Delta \Delta G_{18} \approx \Delta \Delta G_{18}'$) but fails spectacularly to predict the 
difference between the acyl chain lengths $16$ 
and $20$ ($\Delta \Delta G_{20} \ne \Delta \Delta G_{20}'$). The agreement for small mismatch is no surprise since there is reasonable 
agreement between the measured line tension of Powl {\it et al.} (2003) and theory. 
But for 
large mismatch the measured line tension is just far too small to match the data of Perozo  {\it et al.} (2002a). The reader may wonder 
whether this situation might be mitigated by changing the value of $W_O$.   However, it is very difficult to reconcile such small values of 
the line tension with the measured free energy differences of Perozo. 
Perhaps the most distinct characteristic of the data of East, Lee and coworkers is the variation in the line tension 
for large mismatch between proteins and even between Eco and Tb MscL. This would seem to suggest, as they have 
speculated (Powl {\it et al.}, 2003) that conformational changes in the protein (iii) are the most attractive explanation for 
large mismatch dependence of the line tension. As we have already speculated, we expect the static conformation approximation to break
down for large mismatch. It is also difficult to rule out the hypothesis that for large mismatches, a very significant fraction of the 
deformation energy  is not localized in the first ring of lipids surrounding the protein and hence,
is not revealed in the measurements of Powl {\it et al.}. 
A much more meaningful comparison to the data of Perozo {\it et al.} (2002a) might be attempted if the 
same measurements were repeated for the MscL channel trapped in the open state (perhaps via crosslinking). 
This direct measurement of the bilayer interaction free energy would be a useful addition to the experimental story and provide a
direct experimental test of our predicted value of the open channel thickness, $W_O$. 

In our previous paper (Wiggins and Phillips, 2004)  we proposed that the width of the 
hydrophobic region of the protein could be engineered to adjust the 
gating tension of the channel. Shortly after our paper appeared, Yoshimura {\it et al.} (2004) published  data
describing precisely this type of experiment. Yoshimura and coworkers mutated residues in the hydrophobic region of the protein to 
hydrophilic asparagine and studied the gain/loss of function in the mutants. 
Single mutations were shown to possess  significant 
loss of function phenotypes especially for mutations at the boundaries of the hydrophobic interface region of the channel. 
Yoshimura and coworkers also measured the the relative increase in gating pressure which is roughly proportional to the ratio of 
the opening free energies (see eqn \ref{Peqn}).
Of the mutated channels which Yoshimura and coworkers were able to gate, there were mutations which gated at 1.5 times the wild-type
pressure. The most severe loss of function mutations did not gate up to pressure of roughly twice the wild type gating pressure. 
Theoretically, we can estimate the change in the opening free energy due to  these alterations in the protein-lipid interface. 
For a small change in the hydrophobic width of the channel ($dW=dW_O=dW_C$):
\begin{equation}
d \Delta G = -(\Delta F_U)\,d W \approx -3.5 \left(\frac{dW}{1\ \rm \AA}\right)\ kT,
\end{equation} 
for typical values ($W_O=36$ \AA, $W_C=37.5$ \AA, and $n=18$). (Since these patch-clamp measurements were performed in 
spheroplasts rather than synthetic liposomes, the effective lipid parameters are unknown.) 
We expect the change in the opening tension to be roughly
\begin{equation}
\Delta \alpha_{1/2} = \frac{d \Delta G}{\Delta G_A}\,  \alpha_{1/2} \approx 0.3 \left(\frac{dW}{1\ \rm \AA}\right)\ \alpha_{1/2}, 
\end{equation}
where we used the same parameters as above to estimate the relative change in the opening tension. (Remember that the relative change
in the opening tension and pressure will be the same if the patch radius is roughly constant.) 
The free energy changes corresponding to reducing the size of the hydrophobic interface of the protein by a few Angstrom might 
energetically account for the observed increase in gating pressure and perhaps for those channels which did not gate. 
We hope to see this experiment repeated 
in synthetic liposomes where we would have more theoretical control of the system or alternatively studied with the methods 
employed by Powl {\it et al.} (2003) so that the change in the line tension for the closed state might be measured. 

Computationally, thickness deformation of the membrane has been observed in molecular dynamics simulations performed 
by Elmore and Dougherty (2003). Their simulations of MscL in the closed conformation for lipid acyl chain lengths 
10-18 reveal that the lipids at the interface  deform to offset the mismatch, at least {\it in silico}. 
Their simulations have also captured a complementary reduction in the protein hydrophobic interface thickness, a 
conformational change which violates our static conformation approximation 
(as well as the implicit static conformation approximation in Perozo {\it et al}, 2002b,
or Sukharev {\it et al.}, 2001) 
but which we have speculated may play a role in the discrepancy between our theoretical predictions and experimental measurements. 
This protein deformation could, in principle, be 
used to further generalize our analytic model, replacing the static conformation approximation with a model allowing protein 
deformation induced by the membrane, though 
the effective spring constant penalizing lipid-induced protein shape changes  
would need to be determined computationally.   
In fact, the spring constant for the closed state could be 
deduced from the data already provided by Elmore and Dougherty (2003).
This more general model would be a natural extension to the model discussed here.

\subsection{Spontaneous Curvature and Gating}
While we have discussed several quantitative studies of acyl 
chain length versus free energy, the effects of 
spontaneous curvature on gating has, to our knowledge, only been studied by Perozo
{\it et al.} (2002a,b). Perozo and coworkers have shown that
bilayers asymmetrically doped with LPC, a spontaneous-curvature-inducing surfactant, 
can stabilize the open state in the absence of 
tension. In our recent paper (Wiggins and Phillips, 2004) we showed
that  spontaneous-curvature-induced
 line tension could result in precisely this effect.  However, 
we have been unable to make a quantitative analysis of this
idea  since the opening free energy has not been 
measured as a function of LPC concentration. (See 
predictions in figure \ref{gc}.) We expect the concentration dependence of the free energy difference
to be linear in LPC concentration. More complicated scenarios are also possible. If the interface 
tilt is induced by LPC, we would expect a roughly  quadratic rather than linear dependence on LPC concentration.

\section{Conclusion}
The MscL channel is an appealing system in which to study lipid-protein interactions since its function
is to couple tension in the lipid membrane to protein conformation. During the gating transition, the channel undergoes a
very large conformation change, dramatically expanding radially and leading to a significant 
local rearrangement of the lipid bilayer. The deformation free energies induced by this rearrangement
and their role in channel gating has been the focus of this paper. 
While many uncertainties remain, we believe the start of a consistent story has begun to emerge from 
experiment.  Indeed, we speculate that the framework described here might prove useful in analyzing the
function of any ion channel whose gating leads to perturbations in the surrounding membrane.

Our goal in this paper has been to build an analytic framework in which to provide quantitative interpretation and compare  
experimental results on MscL gating.
To that end, we have expanded and improved upon an existing simple analytic membrane-protein model and 
applied it to mechanotransduction and the MscL system. 
%In a recent paper (Wiggins and Phillips, 2004), we used 
%a simplified version of this model to discuss the role of membrane-protein interactions in 
%mechanotransduction.  
In section \ref{FEEaPI}, we have estimated the size of various contributions to the deformation energy of
the membrane and have discussed the scaling of these contributions.
In section \ref{ATMG}, we have shown how this model, when coupled with a simple two state 
static conformational model of the MscL channel qualitatively and quantitatively agrees with most all of the 
experimental features of channel gating, although 
one important geometrical parameter, the open state thickness of the protein, must still be fit.
As part of our analysis, the model suggests that the assumption that protein conformational states 
are independent of their lipid context (such as the lengths of the lipids that the channel is reconstituted in)
is perhaps not borne out experimentally, making structural models of gating even more subtle.
Beyond the interpretation of existing experimental data, we have proposed a number 
of experiments  which we believe will further elucidate the mechanisms of channel gating.
Specifically, additional experiments analogous to those performed by Powl {\it et al.} (2003) with the
channel locked into the open state could provide
topical information about the conformation of the open state and its interaction with the membrane.
Such data,   when combined with the data already available for the closed state, would allow a direct comparison 
to the gating free energies measured by Perozo {\it et al.} (2002a) and a test of our predictions of how the
free energy depends on the geometry of the open state. We still believe that a more controlled version of
the experiments preformed by Yoshimura {\it et al.} (2004), when combined with careful modeling would allow
the sensitivity of the channel to be tuned by changing the size of the hydrophobic interface. We hope that 
these experiments will be repeated in synthetic liposomes where the theoretical model is easier to apply.
Finally, we suggest the need for  a detailed test of the static conformation approximation by a careful measurement of
the area change between states as a function of both applied tension and acyl chain length. We hope that the 
approximations developed in this paper will be useful in precisely formulating  quantitative experimental 
questions.

\section{Acknowledgments}
We are grateful to Doug Rees, Tom Powers, Jan\'e Kondev, Klaus Schulten, Sergei Sukharev and Evan Evans for useful 
discussions, suggestions, and corrections. We also acknowledge support from
NSF grant CMS-0301657 and the NSF funded Center for Integrative Multiscale Modeling and Simulation as
well as the Keck Foundation. 
PAW acknowledges support through an NSF fellowship.

\appendix   
\section{Appendix}
\subsection{Units and Conversions}
Throughout the paper, we  use $kT$ at $T=300\, {\rm K}$ as our energy scale and \AA  as our
fundamental length scale. Tension is in units of 
${\rm pN\ \AA}^{-1}$. This table provides the conversion to ``real life'' units: 
\begin{eqnarray}
T &=& 300\ {\rm K} \\
1\,kT &=& 4.143\times 10^{-14}\,{\rm erg} = 4.143\times 10^{-21}\,{\rm J} = 0.5988\, {\rm kcal\ 
}{\rm mol}^{-1} \\
1\, kT\, {\rm \AA}^{-1} &=& 41.43\ {\rm pN} = 4.143\times 10^{-11}\,{\rm N} \\ 
1\,kT\,{\rm \AA}^{-2} &=& 41.43\ {\rm pN}\ {\rm \AA}^{-1} = 4.143\times 10^{-1}\,
{\rm N\, m}^{-1} \\
1\,kT\,{\rm \AA}^{-3} &=& 4.143\times 10^{10}\, {\rm dyne}\ {\rm cm}^{-2} = 4.143\times 10^{9}\,
{\rm Pa} = 3.0570\times 10^7\,{\rm mmHg} 
\end{eqnarray}

\subsection{Bilayer Parameter Model}
We recommend Rawicz {\it et al.} (2002) (from which we have taken the table below) 
for a brief review of the mechanics of
bilayers. There is a subtlety which we haven't discussed in the paper relating to the 
difference between peak to peak head group thickness which is measured from x-ray
crystallography and mechanical thickness--that portion of the tail of the lipid which is deformed. 
The mechanical thickness of the bilayer is roughly $1$ nm less than the peak
to peak thickness (Rawicz {\it et al.}, 2000). When we discuss the scaling of the moduli, it is this thickness that we really 
consider. This uncertainty is compounded by the question of how
this thickness relates to the hydrophobic thickness of the bilayer. The thickness of the interface 
between the hydrophobic region and the hydrophilic region is also difficult to define
(White and Wimley, 1999). The MscL protein itself does not really have a sudden transition between 
hydrophobic residues and 
hydrophilic ones meaning that one cannot really start with the structure and say definitively 
what the hydrophobic thickness is. This model is at best a caricature which attempts to
capture the essential mechanics and it is for this reason we have not tried to differentiate 
between all these different thickness and replaced them all with a single approximation. 

\begin{figure}
\begin{center}
\epsfig{file=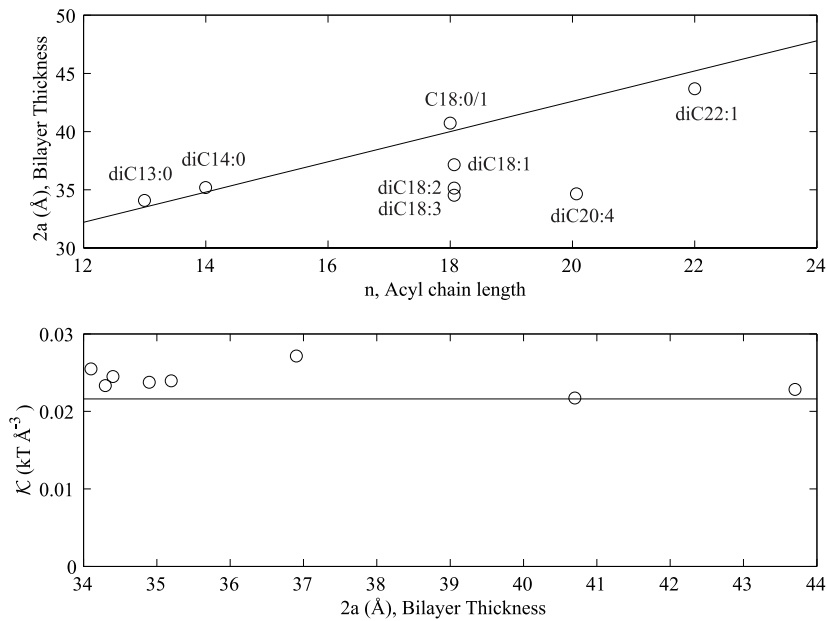}
\caption{ \label{ka} 
Accuracy of lipid model. In the top panel, we plot lipid bilayer thickness versus acyl 
chain width. There is reasonable agreement between the linear fit and the data provided 
that the lipid is not poly-unsaturated. In the bottom panel the effective spring constant 
$\cal K$ is plotted versus lipid bilayer width. $\cal K$ is roughly independent of the 
bilayer thickness. All data from Rawicz {\it et al.} (2000).
}
\end{center}
\end{figure}

Having taken this spartan view of the bilayer, we assume the bilayer acts as if there were 
only one elastic constant governing its behavior, the (effective) Young's Modulus
of the lipid tails:
\begin{equation}
{\cal E} = \epsilon \frac{\Delta V}{V}
\end{equation}
where $\cal E$ is the elastic energy density,  $V$ is the volume and $\epsilon$ is the Young's 
modulus. The only length scale for the bilayer is its thickness $2a$ so all the
rest of the elastic moduli for the bilayer scale with $\epsilon$ and the number of powers of $a$ 
required to get the right units.
These dimensional analysis arguments predict
\begin{eqnarray}
K_B &\propto& a^3, \\
K_A &\propto& a^1, \\
{\cal K} &\propto& a^0.
\end{eqnarray}
This is a rough scaling, not a physical law, but it is sufficient for our calculations. (See figure \ref{ka}.)
For a more 
rigorous argument and experimental results, see Rawicz et. al., 2000. The
table below gives the measured values for the elastic constants of a typical bilayer taken from Rawicz: 
\begin{center}
\begin{tabular}{|ccccc|}
\hline
& Length & $2a$ & $K_A$ & $K_B$ \\
Lipid & (atoms)& \AA & $({kT}/{\rm \AA}^2)$ & $(kT)$ \\
\hline 
C18:0/1* & 18 & $40.7 \pm 0.6$ & $0.568 \pm 0.03$ & $21 \pm 2 $ \\
\hline 
\end{tabular}
\end{center}
$kT$ for $T = 300\ {\rm K}$. Tail Length is the number of carbon atoms which
comprise each of the two tails.   C18:0/1 is 1-oleoyl-2-stearoyl-sn-glycero-3-phosphocholine,

For estimates of bilayer thickness as a function of acyl chain length, we have fit the peak to peak 
head group separation to acyl chain length for the saturated lipids above as shown in figure \ref{ka}.
We have used the relation:
\begin{equation}
2a = 1.3n+16.6\ {\rm \AA},
\end{equation}
although slightly more elaborate formulas are offered in Rawicz {\it et al.} (2000).
When discussing the lipids used by other authors, we have used the same naming convention they employed:
PC12 (12:0 dilauroyl-phosphatidylcholine), PC10 (10:0 dicaproyl-phosphatidylcholine), PC16 
(16:1dipalmitoleoyl-phosphatidylcholine), PC18 (18:1
dioleoyl-phosphatidylcholine), PC20 (20:1 Eicossenoyl-phosphatidylcholine), PE (18:1 dioleoyl-phosphatidylethanolamine), 
LPL (lysophospholipid), 
LPC (lysophosphatidylcholine), 
DOPC (dioleoylphosphatidylcholine), and DOPE (dioleoylphosphatidylethanolamine).

\subsection{Effective free energy density}
%We consider inclusion induced perturbations around a spherical background geometry. We assume a solutions 
%of the form
%\begin{equation}
%r = R_0\left[1+\epsilon\left(\theta,\phi\right)\right],
%\end{equation}
%where the surface is parameterized in the standard spherical coordinates $\theta$ and $\phi$ and 
%we assume $\epsilon\ll 1.$ To lowest order, we can show that the twice the mean curvature takes the
%form:
%\begin{equation}
%{\rm tr}\ {\bf S}(\theta, \phi) = -\frac{2}{R_0} + R_0 \nabla_{S_2}^2 \epsilon(\theta, \phi), 
%\end{equation} 
%where $\nabla_{S_2}^2$ is the Laplacian on the two sphere. By defining $r \equiv R_0 \theta$ and
%$\tilde h(r,\phi) \equiv R_0 \epsilon(r/R_0,\phi)$, and again keeping only the lowest order terms, 
%we get the expected form:
%\begin{equation}
%{\rm tr}\ {\bf S}(\theta, \phi) = -\frac{2}{R_0} + \nabla^2 \tilde h(r, \phi), 
%\end{equation} 
%where $\nabla^2$ is the usual Laplacian. For $r\ll R_0$, the relation between the radial slope
%of the mid-plane and the perturbation is:
%\begin{equation}
%h'(r,\phi) = -\frac{r}{R_0} + \tilde h'(r,\phi).
%\end{equation}
%Since $u$ is defined as a difference, $u(r,\phi) = \tilde u(r,\phi)$.

The mean curvature contributions to the free energy density are: 
\begin{equation}
{\cal G}_{\rm B} = {\textstyle\frac{K_B}{2}}\left[\underbrace{\left(\nabla^2  h\right)^2+
\left(\nabla^2 u\right)^2}_{\cal M}-\underbrace{2\left(C \nabla^2  h +{\overline C}\nabla^2 u\right)}_{\partial \cal M}\right],
\end{equation}
where the variation of the ${\cal M}$ terms contribute to the action in the bulk (bilayer), 
the $\partial{\cal M}$ terms are total derivatives and can be evaluated at the
interface, and constant terms are dropped. The Gaussian curvature contributes only at the 
boundary, not in the bulk, and will be calculated exactly below. 
The tension contributions to the free energy density are
\begin{equation}
{\cal G}_{\alpha} = \underbrace{\frac{\alpha}{2}\left[\left(\nabla  h\right)^2+
\left(\nabla u\right)^2\right]}_{\cal M}
\end{equation}
where as before, the variation of the $\cal M$ terms contribute to the action in the bulk (bilayer). 
The interaction free energy density is 
\begin{equation}
{\cal G}_{\rm I}= \underbrace{\frac{K_A}{2a^2}u^2}_{\cal M}. 
\end{equation} 

\subsection{Equilibrium equations and solutions}
The equations that result from the variation of $u$ and $h$ are
\begin{eqnarray}
0 = \frac{\delta G[u, h]}{\delta u} &=& \left[K_B \nabla^4 - \alpha\nabla^2 +\frac{K_A}{a^2}\right]u, \\
0 = \frac{\delta G[u, h]}{\delta  h} &=& \left[K_B \nabla^4 - \alpha\nabla^2\right] h.
\end{eqnarray}
One Laplacian can be dropped from the equation for $ h$ and amounts  to the freedom for 
rotations of the bilayer in the $x,z$ and $y,z$ planes and displacements
along the z axis. We choose a configuration by specifying the equilibrium position of the plane.
Assuming cylindrical symmetry, these equations are satisfied by the modified Bessel function $K_0$
\begin{equation}
{\textstyle\frac{1}{r}}\partial_r r \partial_r K_0(\beta r) = \beta^2 K_0(\beta r),
\end{equation}
resulting in the solutions
\begin{eqnarray}
u(r) &=& A_+K_0(\beta_+r)+A_-K_0(\beta_-r), \\ 
 h(r) &=& BK_0(\beta_H r),
\end{eqnarray}
where $\beta_\pm$ and $\beta_H$ are given by:
\begin{eqnarray}
\beta_{\pm} &\equiv& \sqrt{ \frac{\alpha\pm\sqrt{\alpha^2-4K_B K_A/a^2}}{2K_B}}, \\
\beta_H &\equiv& \sqrt\frac{\alpha}{K_B},
\end{eqnarray}
where the branch cuts for the square roots are along the negative real axis. $\beta_\pm$ need not be 
real and in fact if
\begin{equation}
\alpha^2<4K_B K_A/a^2,
\end{equation}
the $\beta_\pm$ are complex and $u$ oscillates as it decays. 
The boundary conditions can be used to determine the constants $A_\pm$ and $B$ as
\begin{eqnarray}
A_\pm &=& -\frac{ K_\mp U' + \beta_\mp U K_\mp '}{\beta_\pm K_\pm' K_\mp - \beta_{\mp} K_\pm K_\mp '}, \\
B &=& \frac{H'}{\beta_H K_0'(\beta_H R)},
\end{eqnarray}
where
\begin{eqnarray*}
K_\pm &\equiv& K_0(\beta_\pm R), \\
K_\pm '&\equiv& K_0'(\beta_\pm R). 
\end{eqnarray*}
For large $z$, the Bessel functions can be replaced by their asymptotic approximation as
\begin{eqnarray}
K_0(z) &\rightarrow& \sqrt \frac{\pi}{2z} \exp(-z) \\
K'_0(z) &\rightarrow& -\left(1+\frac{1}{2z}\right)\sqrt{\frac{\pi}{2z}} \exp(-z).
\end{eqnarray}
The relevant length scale for $u$ is the decay length for thickness deformation:
\begin{equation}
\beta^{-1} \equiv \left(\frac{K_A}{K_B a^2}\right)^{-1/4} \sim 11\ {\rm \AA} < R_{\rm MscL}.
\end{equation}
Since the decay length is shorter than the channel radius, we can expand our results in $\beta R$.
By way of  contrast, the length scale for mid-plane deformation is typically much larger since 
the restoring force, in the form of the tension, is relatively weak
\begin{equation}
\beta_H^{-1} = \sqrt{\frac{K_B}{\alpha}} \sim 27\ \sqrt{\frac{\alpha_*}{\alpha}}\ {\rm \AA}.
\end{equation} 
At low tension the length scale is even larger. Fortunately we will see that when the analytic 
approximation breaks down, the mid-plane energy is irrelevant in comparison with the other 
contributions anyway.

\subsection{ Calculation of Free Energy}
\begin{figure}
\begin{center}
\epsfig{file=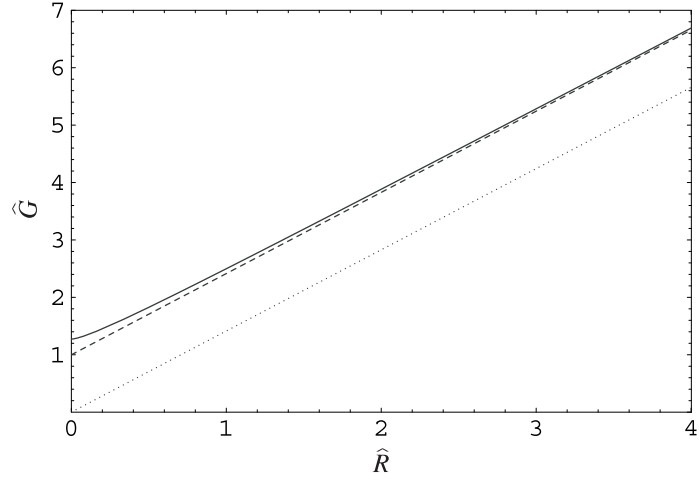}
\caption{\label{thickness} 
Validity of asymptotic approximation for dimensionless thickness deformation free energy. The 
curves above depict the difference between the exact result (eqn. \ref{exactthickness}, solid 
curve), the asymptotic expansion (eqn. \ref{approxthickness}, dashed curve), and the dominant 
scaling result (shown in table \ref{table}, dotted curve). There is excellent agreement between 
the approximate result and exact result for radii relevant for MscL: $\hat R > 1$. }
\end{center}
\end{figure}

Except for the areal deformation term, the free energy can be calculated on the boundary by 
integrating by parts:
\begin{eqnarray}
G[ h,u] &=& \int_{\cal M'} d^2x {\ \cal G}, \\
&=& \int_{\cal M'} d^2x \left(u\frac{\delta G}{\delta u}+ h\frac{\delta G}{\delta 
 h}+\alpha\right)+\int_{\partial \cal M'}ds\ \hat n\cdot \left(...\right), 
\end{eqnarray}
where the variations in the integral over the bilayer ${\cal M}$ are zero since the 
equations for equilibrium are satisfied. The surface integrals come from integration by
parts. The spontaneous and background curvature contributions are
\begin{eqnarray}
G_C &\equiv& -\int_{\cal M'}d^2x\ \left(C_+\nabla^2 h_+-C_-\nabla^2  h_-\right),\\
&=& -\oint_{\partial \cal M'} d\hat n\ \cdot {\textstyle\frac{K_B}{2}}\left(C_+\nabla  h_+-C_-\nabla  h_-\right),\\
&=& \pi R K_B \left(C_+H'_+-C_-H'_-\right). 
\end{eqnarray} 
The energy contributions from thickness deformations of the bilayer are
\begin{eqnarray}
G_{U} &=& \frac{1}{2} \oint_{\partial \cal M'} d\hat n \cdot \left( K_B\left[ \nabla u 
\nabla^2 u- u \nabla^3 u\right] + \alpha u\nabla u \right), \\
&=& \pi R ({-\hat r}) \cdot \left( K_B \left[\nabla u \nabla^2 u -u \nabla^3 u\right] + 
\alpha u\nabla u \right)|_{R}, \\
&=& \pi R K_B \frac{\left(\beta_+^2-\beta_-^2\right)\left(K_+U' - \beta_+ U K_+'\right)
\left(K_-U'-\beta_-UK_-'\right)}{\beta_-K_+K_-'-\beta_+K_-K_+'} - \alpha UU', 
\label{exactthickness}\\
&=& \pi R \left[ K_B\left(\beta_++\beta_-\right)\left(U'+ \left[\beta_++{\textstyle
\frac{1}{2R}}\right] U\right)\left(U' + \left[\beta_-+{\textstyle\frac{1}{2R}}\right] 
U\right)-\alpha UU' \right],
\end{eqnarray}
where $G_U$ contains all the free energy terms in $u$ except those proportional to $C_\pm$ 
and the asymptotic approximation has been used in the last line of the derivation. Consider 
the simple limit when $U'=0$, namely,
\begin{equation}
G_U = \pi K_B \left(\beta_++\beta_-\right)\left[ \beta_+\beta_-R+\left(\beta_++\beta_-
\right)\right]U^2 \label{approxthickness}
\end{equation}
where we have discarded terms in lower powers of $R$. To address the validity of this 
approximation, we compare this result with the exact result. We make the radius 
dimensionless using the inverse decay length in the low tension limit
\begin{eqnarray}
\beta &\equiv& \left(\frac{K_A}{K_Ba^2}\right)^{1/4},\\
\hat{R} &\equiv& \beta R.
\end{eqnarray} 
We define a dimensionless thickness deformation free energy:
\begin{equation}
G_U = \pi K_B U^2 \beta^2 \hat G_U.
\end{equation}
The exact result and the approximation are compared in figure \ref{thickness}.

\begin{figure}
\begin{center}
\epsfig{file=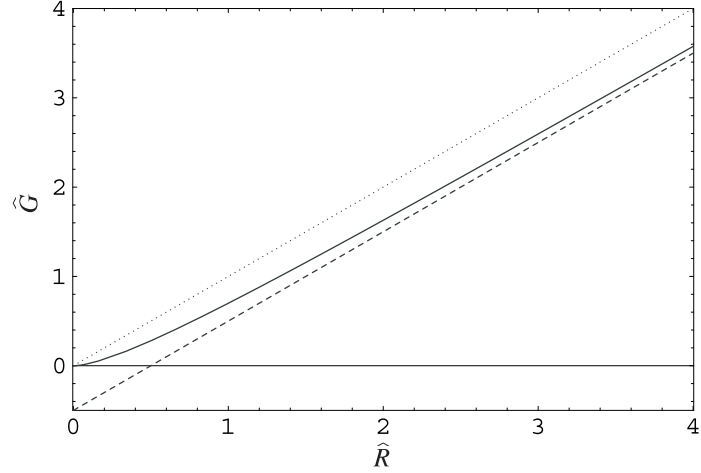}
\caption{\label{midplane} 
Validity of asymptotic approximation for dimensionless mid-plane deformation free energy (eqn \ref{dimless}). 
The curves above depict the difference between the exact result (eqn. \ref{exactmidplane}, 
solid curve), the asymptotic expansion (eqn. \ref{approxmidplane}, dashed curve), and the 
dominant scaling result (table \ref{table}, dotted line). For MscL the prefactor $\pi K_B 
 H'^2$ is typically less than $kT$ implying that the greatest error (when the tension 
is 0) is a fraction of a $kT$ at most. } 
\end{center}
\end{figure}

The free energy associated with the deformation of the mid-plane is
\label{gha}
\begin{eqnarray}
G_{H} &=& \frac{1}{2} \oint_{\partial \cal M'} d\hat n \cdot \left(K_B\left[ \nabla  h 
\nabla^2  h- h \nabla^3  h \right]+ \alpha  h \nabla  h \right), \\
&=& \pi R (-{\hat r}) \cdot \left[\alpha  h\nabla  h \right]_{R}, \\
&=& \pi K_B  H'^2 \hat R \left[{\textstyle \frac{K_0}{|K'_0|}}\right]_{\hat R} \label{exactmidplane} 
\end{eqnarray}
where $\hat R\equiv \beta_H R$. The last line is the exact result of the model. 
If we apply the asymptotic approximation, the result reduces to
\begin{equation}
G_{H} = \pi K_B  H'^2 \left[\hat R-{\textstyle\frac{1}{2}}+{\cal O}({\textstyle\frac{1}
{\hat R}})\right]. \label{approxmidplane}
\end{equation}
The asymptotic approximation is violated for small tensions but the result is typically 
acceptable since the relative error in the energy when the tension is small is irrelevant. 
The prefactor is typically less than a $kT$ and as can be seen in fig. \ref{midplane} the 
error is at most half this prefactor. We define a dimensionless mid-plane deformation 
free energy
\begin{equation}
G_H = \pi K_B  H'^2 \hat G_H. \label{dimless}  
\end{equation}
The dimensionless energy defined above is compared with the approximate value in figure \ref{midplane}.

The Gaussian curvature contribution can be calculated exactly and has no local effect 
because it is related to a well known topological invariant, the Euler Characteristic:
\begin{equation}
2\pi \chi \equiv \int_{\cal M} d^2\sigma\, \det {\bf S}  -  \int_{\partial \cal M} ds\, k,
\end{equation}
where $\det {\bf S}$ is the Gaussian curvature and 
\begin{equation}
k \equiv t^an_b\nabla_a t^b,
\end{equation}
is the curvature of the boundary where $\vec t$ is the unit tangent on the boundary and $\vec n$ 
is the outward pointing unit normal to the boundary. See Polchinski, 1998,  for example. In terms of the 
Euler Characteristic, the Gaussian curvature contribution is
\begin{equation}
G_G = K_G\left(2 \pi \chi+  \int_{\partial \cal M} ds\, k\right).
\end{equation}
$\chi$ depends on membrane topology alone and can be dropped since changes in protein conformation do not effect the membrane topology.
The Gaussian bending energy is therefore exactly
\begin{equation}
G_G = K_G 2 \pi \cos \theta,
\end{equation}
where $\theta$ is the angle of the bilayer away from horizontal at the interface. For the bilayer model in the small angle approximation this is
\begin{equation}
G_G = -\frac{K_G \pi}{2} \left(H_+'^{\, 2}+H_-'^{\,2}\right) =  -K_G \pi \left(H'^{\, 2}+U'^{\, 2}\right).
\end{equation} 
The Gaussian curvature contribution induces bending of the protein to relieve the bending of the bilayer. Existing measurements are consistent with \begin{equation}
K_G<-{\textstyle\frac{1}{2}}K_B,
\end{equation}
(see Boal, 2002) for references) but we will assume that the magnitudes are similar. If this is the case, none of these corrections is particularly relevant for MscL.

\label{tensionexp} 
Finally we calculate the areal deformation term. Before explaining the calculation, let us define precisely what we mean by the tension. The tension we are discussing is the applied tension, not a surface tension. Changes in the inclusion conformation do not effect the area of the bilayer--it is assumed that there is some small change in the global conformation which absorbs this area change. Furthermore these conformational changes do not change the tension since we assume that the bilayer is much larger than the size of the inclusion. Since the area of the bilayer is essentially fixed--at least the number of lipid molecules in the bilayer is fixed--the tension we discuss here is the applied tension rather than a surface tension. 

The global conformation of the bilayer acts as a bilayer reservoir. The free energy cost for increasing the bilayer area of our small system is:
\begin{equation}
dG_A = \alpha dA_{\cal M} = -\alpha dA_{P}
\end{equation}
where the change in the proteins area is minus that of the bilayers. As mentioned above we assume that the reservoir is large enough that changes in the protein conformation have no effect on the tension.

\subsection{ Saturation of Thickness Deformation }
If the mismatch $2|U|$ is less than $2U_*$, then the mismatch is entirely absorbed by thickness deformation.
The maximum thickness deformation free energy, corresponding to a mismatch of $2U_*$, is
\begin{equation} 
G_U^{\rm Max} = \frac{4 \pi R \sigma_*^2}{{\cal K}\left(1+\frac{\sqrt{2}}{\beta R}\right)} = 14\, kT,
\end{equation}
evaluated for the closed state.
For larger mismatches, $2U_*$ is absorbed by the thickness deformation while $2(|U|-U*)$ is exposed to the 
solvent. The combined interface and thickness deformation free energy for $|U|>U_*$ is
\begin{equation}
G_{UW} = \sigma_* 2\pi R \left(2|U|-U_*\right).
\end{equation}
This correction does not dramatically effect the qualitative picture of the thickness deformation discussed 
above. In fact, in fig. \ref{saturation_data} we have plotted the deformation energies for interface energy alone, thickness 
deformation alone, and the corrected thickness deformation to show that for the range of bilayer widths 
of interest in this problem, there is little difference between thickness deformation and the corrected 
thickness deformation, while ignoring thickness deformation altogether in favor of interface energy alone 
results in a significant error. 
 
\subsection{Details of the Perozo vs Powl comparison}
Below we have estimated the bilayer deformation energy based on the EcoMscL data of Powl {\it et al.} (2003). For the
closed states, we have simply used the values measured by Powl and coworkers. For the open state, we have used a fixed
value of $W_O$ listed below and estimated the mismatch. From the mismatch, we have reinterpreted the data of Powl {\it et al.} 
(2003) as a function of mismatch (see figure \ref{sigma_n_powl}) to estimate the line tension. From the line tensions, we
then compute the deformation energy difference ($\Delta G\equiv 2\pi\left[f_O R_O-f_CR_C\right] $).  
For $W_O=28.0$ \AA:
\begin{center}
\begin{tabular}{c|cccccc}
$n$ & $2a\ ({\rm \AA})$ & $U_O\ ({\rm \AA})$ & $U_C\ ({\rm \AA})$ & $f_O\ (kT{\rm nm}^{-1})$ & $f_C\ (kT{\rm nm}^{-1})$ & $\Delta G\ (kT)$ \\ 
\hline
$16$ &   $37.4$         & $4.7$              & $0.0$              & $0.93$                      & $0.0$     &  $20$    \\
$18$ &   $40.0$         & $6.0$              & $1.3$              & $1.1$                       & $0.15$    &  $22$    \\
$20$ &   $42.6$         & $7.3$              & $2.6$              & $1.2$                       & $0.7$     &  $16$
\end{tabular}
\end{center}
For $W_O=36$ \AA:
\begin{center}
\begin{tabular}{c|cccccc}
$n$ & $2a\ ({\rm \AA})$ & $U_O\ ({\rm \AA})$ & $U_C\ ({\rm \AA})$ & $f_O\ (kT{\rm nm}^{-1})$ & $f_C\ (kT{\rm nm}^{-1})$ & $\Delta G\ (kT)$ \\ 
\hline
$16$ &   $37.4$         & $0.7$              & $0.0$              & $0.07$                      & $0.0$     &  $1.5$    \\
$18$ &   $40.0$         & $2.0$              & $1.3$              & $0.4$                       & $0.15$    &  $6.6$    \\
$20$ &   $42.6$         & $3.3$              & $2.6$              & $0.8$                       & $0.7$     &  $7.5$
\end{tabular}
\end{center}

%\subsection{Caveats}
%One might well question the applicability and validity of such a model where the energy contribution comes from such a small localized area: a shell of about $10\ \rm \AA$ in thickness
%around the inclusion corresponding to just a few lipid layers. The numbers we have used in our paper are properties measured at small curvatures on larger scales. Our answer
%to these objections is simply this: we are building an analytic model, not to predict the energies to high accuracy but to understand the scaling and roughly estimate the size of
%the effects. 

%We briefly mentioned in the spontaneous curvature section that in highly curved regions it is possible that the different components of the bilayer might separate. This becomes
%even more of a problem in light of the thickness deformation free energy. For example shorter tailed lipids added to the bilayer might significantly reduce the thickness
%deformation free energy by surrounding a short protein. If we proceed blindly we would conclude that spontaneous curvature is reducing the free energy (and $H'$ is non-zero) rather
%than concluding that this is another manifestation of the thickness deformation free energy. We will return to this issue again when we discuss experiments. 

\end{document}